\documentclass[aps,prb,twocolumn,superscriptaddress]{revtex4-1}
\usepackage{blindtext}
\usepackage{amsmath,amssymb,amsfonts}
\usepackage{bbm}
\usepackage{bm}
\usepackage{color}
\usepackage{epstopdf}
\usepackage[normalem]{ulem} 
\usepackage{soul} 

\usepackage{graphicx}
\usepackage{indentfirst}
\usepackage{psfrag}
\usepackage{epsfig}
\def\ANGST{{{\buildrel _{\circ} \over {\mathrm{A}}}}}

\newcommand{\bra}[1]{\ensuremath{\left\langle#1\right|}}
\newcommand{\ket}[1]{\ensuremath{\left|#1\right\rangle}}

\newcommand{\nbth}{n_b^{\mathrm{th}}}
\newcommand{\nbrad}{{n_{\mathrm{rad}}}}
\newcommand{\neff}{{n_b^{\mathrm{incoh}}}}

\newcommand{\mr}[1]{\ensuremath{\text{#1}}}
\newcommand{\mean}[1]{\langle{#1}\rangle} 
\newcommand{\ud}[1]{{#1^{\dagger}}}

\newcommand{\diag}{\ensuremath{\text{diag}}}

\renewcommand{\Re}{\mathrm{Re}}

\begin{document}
	\title{
		QED Description of Raman Scattering from Molecules in Plasmonic Cavities
	}
	\author{Mikolaj K. Schmidt}
	\email{mikolaj_schmidt@ehu.es}
	\affiliation{Materials Physics Center CSIC-UPV/EHU, Paseo Manuel de Lardizabal 5, 20018 Donostia-San Sebasti{\'a}n, Spain}
	\affiliation{Donostia International Physics Center DIPC, Paseo Manuel de Lardizabal 4, 20018 Donostia-San Sebasti{\'a}n, Spain}
	\author{Ruben Esteban}
	\affiliation{Donostia International Physics Center DIPC, Paseo Manuel de Lardizabal 4, 20018 Donostia-San Sebasti{\'a}n, Spain}
	\author{Alejandro Gonz{\'a}lez-Tudela}
	\affiliation{Max-Planck-Institut f\"ur Quantenoptik, Hans-Kopfermann-Str. 1., 85748 Garching, Germany}
	\author{Geza Giedke}
	\affiliation{Donostia International Physics Center DIPC, Paseo Manuel de Lardizabal 4, 20018 Donostia-San Sebasti{\'a}n, Spain}
	\affiliation{IKERBASQUE, Basque Foundation for Science, Maria Diaz de Haro 3, 48013, Bilbao, Spain}
	\author{Javier Aizpurua}
	\email{aizpurua@ehu.eus}
	\affiliation{Materials Physics Center CSIC-UPV/EHU, Paseo Manuel de Lardizabal 5, 20018 Donostia-San Sebasti{\'a}n, Spain}
	\affiliation{Donostia International Physics Center DIPC, Paseo Manuel de Lardizabal 4, 20018 Donostia-San Sebasti{\'a}n, Spain}

	\begin{abstract}
		Plasmon-enhanced Raman scattering can push single-molecule vibrational spectroscopy beyond a regime addressable by classical electrodynamics. We employ a quantum electrodynamics (QED) description of the coherent interaction of plasmons and molecular vibrations that reveal the emergence of nonlinearities in the inelastic response of the system. For realistic situations, we predict the onset of \textit{phonon-stimulated Raman scattering} and a counterintuitive dependence of the anti-Stokes emission on the frequency of excitation. We further show that this QED framework opens a venue to analyze the correlations of photons emitted from a plasmonic cavity.
	\end{abstract}		
	\maketitle
	
	Surface Enhanced Raman Scattering (SERS) is a spectroscopic technique in which the inelastic scattering from a molecule is increased by placing it in a \textit{hotspot} of a plasmonic cavity, where the electric fields associated with the incident and the scattered photons are strongly enhanced (see the schematic in Fig.~1(a)).\cite{LeRu20091} The difference between the energy of those two photons provides a \textit{fingerprint} of the molecule, \textit{i.e.}, detailed chemical information about its vibrational structure. {Since the  initial observation of Raman scattering from single molecules,\cite{nie97,kneipp1997single} the use of a variety of plasmonic structures that act as effective optical nanoantennas over the last decades has allowed a tremendous advance of this molecular spectroscopy\cite{sharma2012sers}.  Metallic particles such as nanoshells\cite{jackson2004,shiners2010}, nanorings\cite{Crozier2010}, nanorods\cite{murphy2013}, nanowires\cite{pucci2013}, or nano
		stars\cite{niu2015},  as well as plasmonic nanogap structures formed in particle dimers\cite{xu99,talley2005,zhu2014quantum}, nanoparticle-on-a-mirror morphologies\cite{zeehwankim2008,Lorenzo2010,sonntag2014,Lombardi2016}, or nanoclusters\cite{Ye2012}, are among the variety of structures 
		that offer huge and controllable enhancements of the field intensity in their hotspots, boosting the inherently weak Raman scattering intensity,\cite{PhysRevE.62.4318,LeRu20091}
		and ultimately enabling the chemical identification and imaging of particular vibrational modes of a molecule with subnanometer resolution.\cite{zhang2013chemical}} These results suggest that some experiments might have reached the regime where the quantum-mechanical nature of both the molecular vibrations and the plasmonic cavity emerges,\cite{shalabney2015coherent} and call for an adequate theoretical description that goes beyond the classical treatment of the electric fields produced in plasmonic cavities.\cite{LeRu20091,johansson2005surface,PhysRevLett.93.243002}
	
	In this work we address the underlying quantum-mechanical nature of Raman scattering processes by quantizing as bosonic excitations both the vibrations of the molecule and the electromagnetic field of a plasmonic cavity. The description of the vibrations through bosonic operators can be justified by considering the harmonic approximation to the energy landscape of the molecule along a generalized atomic coordinate (Fig.~1(b)), such as the length of a molecular bond, \textit{e.g.}, C$=$O.\cite{johansson2005surface,shalabney2015coherent} These vibrations interact with the cavity photons through a nonlinear Hamiltonian, reminiscent of that found in optomechanical systems.\cite{RevModPhys.86.1391} In this description, the large enhancement of the Raman scattering from a molecule in the plasmonic cavity occurs thanks to the significant shrinking of the effective mode volume of a single photon. On the other hand, such plasmonic systems experience strong Ohmic losses - that is, they are characterized by very low \textit{quality factors} (typically $Q \lesssim 10$) - placing these systems deep in the \textit{bad cavity} regime (see the sketch of a plasmon mode in Fig.~1(c)). This system was recently addressed using a linearization scheme based on the classical Langevin equation for the displacements of plasmon and phonon fields.\cite{roelli2014molecular,schmidt2016nanocavities} 
	
	In this work we complement this approach by employing an exact numerical solution { of the quantum mechanical dynamics of the nonresonant Raman scattering process in a plasmonic cavity} that fully describes the buildup of incoherent population of vibrations in the molecule, addressing the effect of phonon stimulated Stokes scattering, and furthermore, allows to account for the quantum and classical covariances between the two bosonic fields. {This formalism thus predicts a variety of nonlinear quantum effects in the scattered signal that might have already been revealed in different Raman measurements, and opens the door for specific design of experimental configurations that can test, and eventually control, the underlying coherences within SERS.}
	
	\begin{figure}[htbp!]
		\begin{center}
			\includegraphics[width=.9\columnwidth]{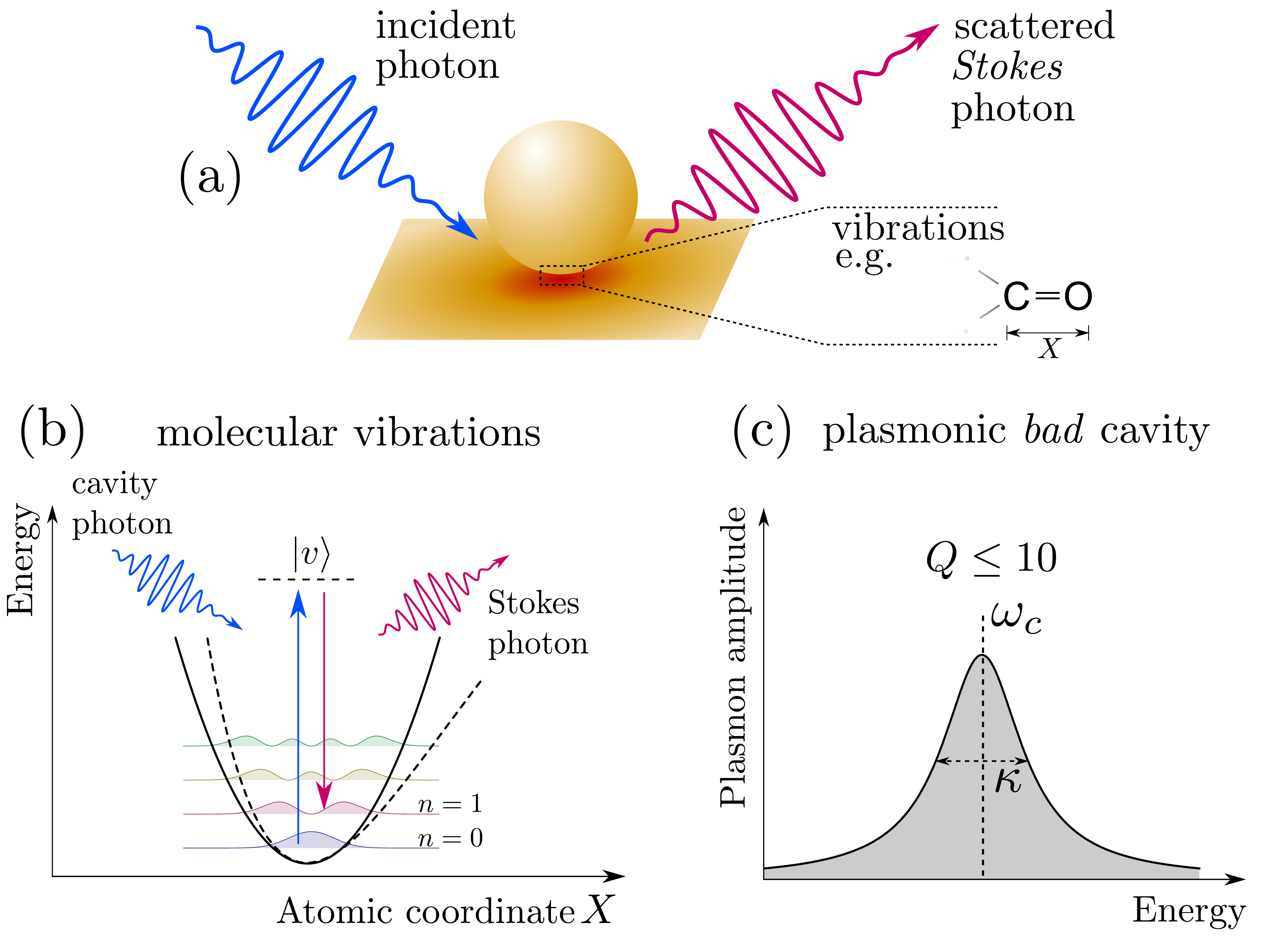}
			\caption{(a) Schematic of a setup for observing the Raman scattering from a molecule placed in a plasmonic cavity. (b) Schematic of the two-photon nonresonant Stokes scattering between two vibrational states of the molecule ($n=0\rightarrow 1$) mediated by a virtual state $\ket{v}$. A harmonic potential (solid lines) approximates the energy landscape of the ground electronic level (dashed lines). (c) Typical plasmonic cavity mode centered at the frequency $\omega_c$ in the visible range, with the width associated with the decay rate $\kappa$ related to $\omega_c$ through the \textit{quality factor} $Q$=$\omega_c/\kappa$.  
			}
		\end{center}
	\end{figure}
	
	\section{Results and Discussion}
	
	\subsection{Theoretical Framework}
	
	{The interaction between plasmons and vibrations that governs the dynamics of a SERS process can be properly addressed by establishing the Hamiltonian of the system.} To derive this Hamiltonian, let us consider the vibrational states of the ground electronic state of the molecule. In a simplified one-dimensional model the potential landscape of this state can be approximated as a displaced harmonic potential. Therefore, the vibrational levels separated by \textit{phonon} frequency $\omega_m$\cite{bougeard2009calculation} can be quantized using the creation and annihilation operators $\hat{b}$ and $\hat{b}^{\dag}$. We can thus define a linear polarizability of the molecule along this coordinate as $\hat{\alpha}_{\nu} = R_{\nu} Q_{\nu}^0(\hat{b}+\hat{b}^{\dag})$,\cite{LeRu20091} where $R_{\nu}$ is the element of the Raman tensor and $Q_{\nu}^0$ is the zero-point amplitude of the vibrations. The molecule is coupled to the quantized field of the cavity mode with resonant frequency $\omega_c$ and effective volume $V_{\text{eff}}$, which can be expressed by the plasmon annihilation ($\hat{a}$) and creation ($\hat{a}^{\dag}$) operators as $\hat{E} = i \sqrt{{\frac{\hbar \omega_c}{2 \varepsilon V_{\text{eff}}}}}(\hat{a}-\hat{a}^{\dag})$, where $\varepsilon$ is the permittivity of the medium. Thus the induced \textit{Raman dipole} $\hat{p}_R = \hat{\alpha}_{\nu} \hat{E}$ will be interacting with the cavity field $\hat{E}$, yielding the interaction Hamiltonian $\hat{H}_I = -\hat{p}_R \hat{E}$, which, after dropping the fast rotating terms $\propto (\hat{a}^{\dag})^2$ and $(\hat{a})^2$, and redefining the equilibrium position of the vibrations, yields the non-linear interaction Hamiltonian ($\hbar=1$): $\hat{H}_I = - g \hat{a}^{\dag} \hat{a} (\hat{b}^{\dag} + \hat{b})$,
	with the coupling coefficient $g=R_{\nu} Q_{\nu}^0 \omega_c/(\varepsilon V_{\text{eff}})$. This definition can be related to the resonant emitter-plasmon coupling parameter of the Jaynes-Cummings Hamiltonian $g_{JC}$ and the Purcell factor $F_P\propto Q/V_{\text{eff}}$ (see the discussion in Supporting Information). Interestingly, the interaction Hamiltonian $\hat{H}_I$ is identical to the one used in optomechanical systems, in which the quantized oscillations of a cavity mirror modify the resonance frequency of the cavity.\cite{RevModPhys.86.1391,roelli2014molecular}
	
	It should be noted that this approach is limited to the \textit{off-resonant Raman scattering}, for which the virtual state mediating the Raman transition (Fig.~1(b)) is strongly detuned from an excited electronic state. Also, our Hamiltonian does not consider Raman processes that involve the effect of the simultaneous excitation or emission of two or more phonons, which may become important for very intense lasers.
	
	Furthermore, our model assumes that the plasmonic system can be approximated as a single cavity mode, and described through the canonical quantization scheme. Although this approach should yield an accurate description for the many of plasmonic SERS setups, a more exact description could be offered by a rigorous quantization method which takes into account the exact Green's function of the system \cite{scheel2008macroscopic,dung2003electromagnetic} and would allow consideration of complex plasmonic responses involving many modes.
	
	The coherent evolution of the system, including an additional driving term {(laser)} at frequency $\omega_l$ is then given by the full Hamiltonian
	\begin{equation}\label{Hamiltonian}
	\hat{H} = \omega_m \hat{b}^{\dag} \hat{b} + \omega_{c} \hat{a}^{\dag} \hat{a} - g \hat{a}^{\dag} \hat{a} (\hat{b}^{\dag} + \hat{b}) + i \Omega  (\hat{a}^{\dag} e^{- i \omega_l t} - \hat{a} e^{i \omega_l t}).
	\end{equation}
	Throughout the paper $\Omega^2$ will be referred to as a \textit{pumping power} proportional to the power density of the input laser and the intrinsic parameters of the cavity mode.
	
	The master equation for the dynamics of the density matrix of the system $\rho$ in the absence of pure dephasing\cite{yampolsky2014seeing} reads
	\begin{equation}\label{master.equation}
	\partial_t \rho  = i[\rho,\hat{H}] + \frac{\kappa}{2} \mathcal{D}_{\hat{a}}[\rho]+ \frac{(\nbth+1)\gamma_m}{2}\mathcal{D}_{\hat{b}}[\rho] +  \frac{\nbth \gamma_m}{2} \mathcal{D}_{{\hat{b}}^{\dag}}[\rho],
	\end{equation}
	where the last three terms on the right-hand side are Lindblad-Kossakowski terms defined as\cite{gorini1976completely, lindblad} $\mathcal{D}_{\hat{O}}[\rho] = 2 \hat{O}\rho\hat{O}^{\dag} - \hat{O}^{\dag} \hat{O}\rho - \rho\hat{O}^{\dag} \hat{O},$
	describing the decay of the plasmons ($\mathcal{D}_{\hat{a}}$), phonons ($\mathcal{D}_{\hat{b}}$), and the thermal pumping ($\mathcal{D}_{{\hat{b}}^{\dag}}$) of the phonons by the environment at temperature $T$, with the thermal population $\nbth=(e^{\hbar \omega_{m}/k_B T}-1)^{-1}$. 
	
	In the following, we take the phonon energy $\omega_m=0.1$ eV and the phonon decay rate $\gamma_{m}=1$~meV,\cite{yampolsky2014seeing} as representative values that characterize vibrational fingerprints, and consider a typical plasmonic cavity with quality factor of $Q=10$ resonant at $\omega_c=2.5$ eV, decaying with rate $\kappa=0.25$ eV (see Supp. Info.). The coupling parameter is taken as $g=1$~meV in accordance with the reported characteristics of {state-of-the-art resonant plasmon-emitter systems (see Supp. Info.)}.
	
	To analyze the dynamics of the system, we follow three complementary approaches. First, we solve directly the master equation in Eq.~(\ref{master.equation}) by means of numerical calculations,\cite{navarrete2015open} (taking advantage of low excitation numbers to truncate the Hilbert space). We complement this numerical solution by two other approximate treatments that allow to obtain analytical results, useful to identify the influence of the different parameters in the Raman signal and thus provide simpler expressions to interpret experimental results. In the first analytical approximation, we linearize the Hamiltonian to a purely quadratic form which applies to the weak coupling ($g\ll\kappa$) regime we are interested in, and allows for the analytical treatment (see Supporting Information for details). We note that when solving the linearized Hamiltonian, we do not map the quantum Langevin equation to the classical dynamics equations, as is often done in the analysis of optomechanical systems.\cite{RevModPhys.86.1391} Consequently, this approach provides a complete characterization of the classical and quantum correlations within the system. Finally, to develop the second analytical approach, we apply the \textit{quantum noise approach},\cite{RevModPhys.82.1155} in which the vibrations of the molecule are coupled to a bath of fluctuating population of the cavity plasmons.\cite{roelli2014molecular} This last formalism allows us to describe the effects of the \textit{static} and \textit{dynamical backaction}, which arise when the vibrations of the molecule modify the resonant frequency of the cavity and, therefore, the population term in the Hamiltonian (Eq.~(\ref{Hamiltonian})).\cite{RevModPhys.86.1391,aspelmeyer2014cavity} The static backaction describes the constant, coherent displacement of the state of the vibrations, and can be largely neglected in the regime of parameters of interest (see the Methods section for a further discussion). On the other hand, the dynamical backaction (DBA), observed when the incident laser is detuned from the cavity, leads to the heating or cooling of the vibrations.\cite{RevModPhys.86.1391} The quantum noise approach captures those effects, at the price of losing the information about coherences in the molecule-plasmon system.\cite{RevModPhys.86.1391,aspelmeyer2014cavity} The spectra of emission from the cavity are calculated following Glauber's photodetection theory,\cite{glauber63a} as $I(\omega)=\alpha_{\text{det}} S(\omega)$, where the frequency-independent parameter $\alpha_{\text{det}}$ describes the properties of the detection system\cite{johansson2005surface} and $S(\omega) = \omega^4 \int_{-\infty}^{\infty} dt e^{- i\omega t}\langle \hat{a}^\dag(t)\hat{a}(0) \rangle_{\text{ss}}$ in the steady state (ss). The two-time correlator is calculated by applying the \textit{quantum regression theorem} (QRT).\cite{carmichael2009statisticalVol1} 
	
	Let us first focus on the numerical solution. Two spectra of emission from the cavity, calculated for the weak ($\Omega^2=10^{-2}$~eV$^2$) and strong illuminations ($\Omega^2=0.5$~eV$^2$) and for the laser tuned to the cavity ($\Delta\equiv\omega_c-\omega_l=0$), are shown in Fig.~2(a) and (b), respectively. In the inset of Fig.~2(a) we zoom in on the anti-Stokes emission calculated for the environment at $T=0$~K (dashed line) and $T=300$~K (solid line; $\nbth(T=300~$K$)\approx0.02$). The difference between these plots illustrates the effect of \textit{thermal pumping} of the vibrational levels by the environment.\cite{PhysRevLett.76.2444,maher2006conclusive}
	
	To further explore how the signal evolves with the pumping power, we plot in Fig.~2(c) the maxima of the Stokes (blue solid line; in the region of interest this function is independent of the temperature $T$) and anti-Stokes (orange dashed line for $T=0$~K and orange solid line for  $T=300$~K) emission lines for the increasing $\Omega^2$. In the weak pumping regime ($\Omega^2 \lesssim 10^{-2}$~eV$^2$) for nonzero temperature (solid lines) both the Stokes $S(\omega_S)$ and anti-Stokes $S(\omega_{aS})$ emission depend linearly on $\Omega^2$, indicating that the anti-Stokes transition originates from the thermally excited vibrational state. For higher driving powers ($10^{-2}$~eV$^2 \lesssim \Omega^2 \lesssim 0.5$~eV$^2$) the anti-Stokes intensities become independent of the temperature, as the phonons are provided primarily by the Stokes transitions (\textit{vibrational pumping}).\cite{maher2006conclusive}
	The transition between the thermal and the vibrational pumping of phonons\cite{maher2006conclusive} is illustrated in Fig.~2(d), where we plot the populations of the phonons (green line) and plasmons (red line) for $T=0$~K (dashes lines) and $T=300$~K (solid lines).
	\begin{figure}[htbp!]
		\begin{center}
			\includegraphics[width=\columnwidth]{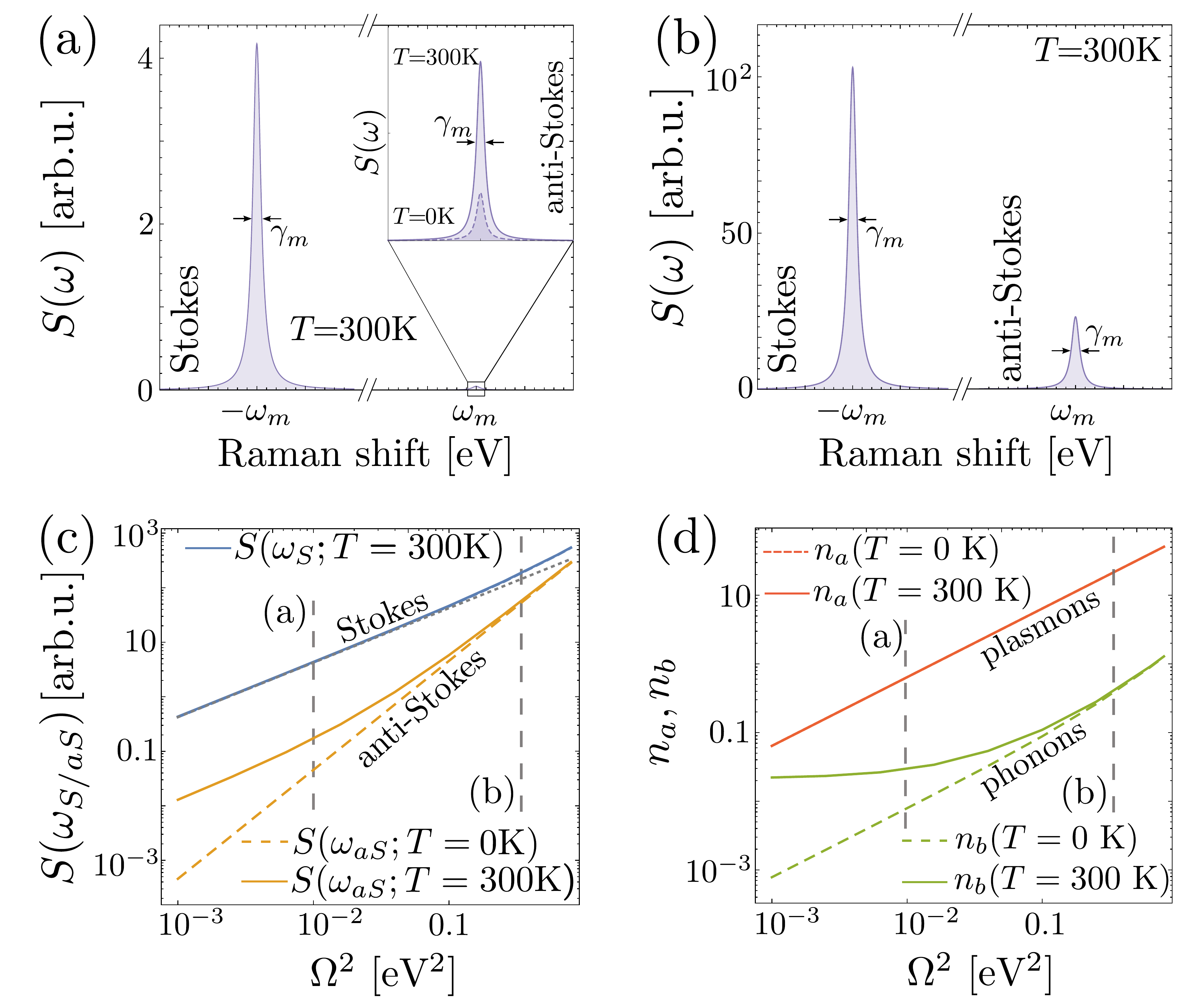}
			\caption{Dependence of the Raman scattering on the excitation power and temperature: (a,b) emission spectra $S(\omega)$ of the molecule in the plasmonic cavity for pumping (a) $\Omega^2=10^{-2}$~eV$^2$ and (b) $\Omega^2=0.5$~eV$^2$ at $T=0$~K (dashed lines) and $T=300$~K (solid lines); 
				(c) emission intensities of the Stokes ($S(\omega_S)$, blue lines, calculated at $T=300$~K) and anti-Stokes photons ($S(\omega_{aS})$, orange dashed and solid lines for $T=0$~K and $T=300$~K, respectively) as a function of the pumping power $\Omega^2$; (d) populations of plasmons (red line) and phonons (green lines) in the steady state for $T=0$~K (dashed lines) or $T=300$~K (solid lines). Vertical dashed lines in (c,d) indicate the pumping powers for plots in (a) and (b). All the cases assume $\Delta=0$.
			}\label{fig2}
		\end{center}
	\end{figure}
	
	Finally, for the highest considered pumping powers ($\Omega^2 \gtrsim 0.5$~eV$^2$) the Stokes intensity $S(\omega_S)$ visibly surpasses the expected linear dependence on $\Omega^2$ (marked with dotted gray line in the top-right corner of Fig.~2(c)). {To address this effect, we develop the first of the aforementioned analytical approaches, by considering the linearized Hamiltonian and employing the QRT which, for the particular case of $\Delta=0, \gamma_m\ll \omega_m,\kappa$ and environment temperature $T$, yields the following expressions for the Stokes and anti-Stokes photons}: 
	\begin{equation}\label{qrtStokes2}
	S(\omega_S) = \frac{4 \omega_S^4}{\gamma_m} s_2 \Omega^2 \left(1+\nbth+s_2 \Omega^2 \frac{\kappa}{\gamma_m}\right),
	\end{equation}
	and
	\begin{equation}\label{qrtantiStokes2}
	S(\omega_{aS}) = \frac{4\omega_{aS}^4}{\gamma_m} s_2 \Omega^2 \left(\nbth + s_2 \Omega^2 \frac{\kappa}{\gamma_m}\right),
	\end{equation}
	with $s_2\approx[4 g/(\kappa |\kappa-2 i \omega_m|)]^2$ (see Supp. Info. for the complete derivation). The first term in the brackets of Eq.~(\ref{qrtStokes2}) can be recognized as the conventional two-photon cavity-assisted Stokes transition, linearly dependent on $\Omega^2$. As we show in the Supp. Info., by considering the enhancement $K=|E/E_0|$ of the electric component of the incident illumination $E_0$ {produced} by the plasmonic cavity, and relating it through the reciprocity theorem to the Purcell factor of the cavity,\cite{Carminati:98,reciprocity} this term in Eq.~(\ref{qrtStokes2}) allows us to retrieve the expected dependence of the Stokes emission $S(\omega_S)\propto K^4$.
	
	The sum of the second and the third terms in brackets in Eq.~(\ref{qrtStokes2}), or the two terms in brackets in Eq.~(\ref{qrtantiStokes2}), represents the incoherent population of the phonon mode ($n_b^{\text{incoh}}=\langle \hat{b}^{\dag}\hat{b}\rangle_{\text{ss}}-\langle \hat{b}^{\dag}\rangle_{\text{ss}}\langle \hat{b}\rangle_{\text{ss}}$) arising from (i) the thermal pumping and (ii) the coupling to the plasmon cavity, respectively. These two terms together describe a process of \emph{phonon-stimulated Raman scattering}, in which the population of vibrations enhances the rate of Stokes scattering. Therefore, the last term in Eq. (\ref{qrtStokes2}) explains the nonlinearity of the Stokes signal for large $\Omega^2$. We note that phonon-stimulated Raman scattering has been reported in experiments on ensembles of Raman-active centers, \textit{e.g.}, hydrogen gas (see Refs.~[\!\!\citenum{Wang,stimulated}] and references therein). However, to our knowledge, the criterion for the onset of this effect for single scatterers in a cavity (see the Supporting Information for an explicit formulation) has not been reported in the literature.
	
	\begin{figure*}[htbp!]
		\begin{center}
			\includegraphics[width=\textwidth]{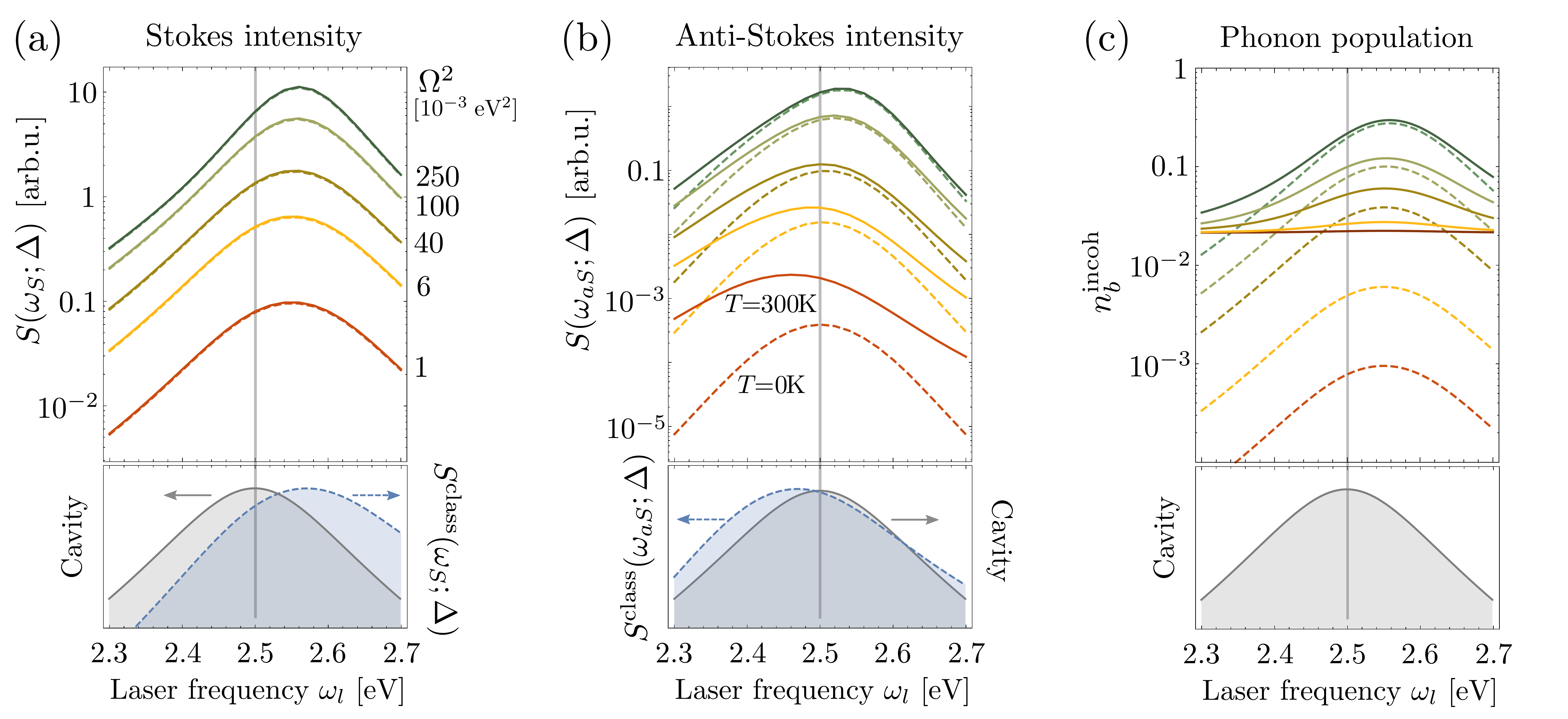}
			\caption{Dependence of the (a) Stokes and (b) anti-Stokes emission and (c) phonon population on the frequency of the incident laser $\omega_l$. In the two left-most top panels we show the numerically calculated intensities  at the (a) Stokes $S(\omega_S;\Delta=\omega_c-\omega_l)$ and (b) anti-Stokes $S(\omega_{aS};\Delta)$ frequencies for the pumping power from $\Omega^2=10^{-3}$~eV (red lines) to $\Omega^2=0.25$~eV (green lines) and environment temperature $T=0$~K (dashed lines) and $T=300$~K (solid lines). In bottom panels in (a, b) the dashed blue lines show the predictions of classical models (Eq.~(\ref{classical})) for (a) $S^{\text{class}}(\omega_{S}; \Delta)$ and (b) $S^{\text{class}}(\omega_{aS}; \Delta)$, whereas the cavity resonances are shown in (a-c) in gray. In (c) we plot the incoherent phonon population $\neff$ calculated within the quantum noise approach [Eq.~(\ref{ntotal})]. Identical results are obtained with the exact numerical calculation. The temperature and intensity for each situation are denoted with the same color code as in (a) and (b).}\label{detuning.fig}
		\end{center}
	\end{figure*}
	
	To further explore the effects of the thermal and vibrational pumping of phonons on Raman scattering, we consider the dependence of the Raman scattering on the detuning $\Delta=\omega_c-\omega_l$ of the incident laser. In the typical classical models of SERS\cite{alonso2012resolving} the dependence of the Stokes ($S^{\text{class}}(\omega_S; \Delta)$) and the anti-Stokes ($S^{\text{class}}(\omega_{aS}; \Delta)$) emission is determined by the enhancement of the electric field of both the incoming ($|E(\omega_l)/E_0(\omega_l)|^2$) and outgoing ($|E(\omega_{S/aS})/E_0(\omega_{S/aS})|^2$) photons at the position of the molecule:
	\begin{equation}\label{classical}
	S^{\text{class}}(\omega_{S/aS}; \Delta) \propto \omega_{S/aS}^4\left|\frac{E(\omega_l)}{E_0(\omega_l)}\frac{E(\omega_{S/aS})}{E_0(\omega_{S/aS})}\right|^2.
	\end{equation}
	Assuming that the enhancement is given by a Lorentzian profile centered on the cavity resonance $\omega_c$ with width $\kappa$ (gray curves in the bottom panels of Fig.~3), we expect that the calculated Raman emission spectra $S(\omega_{S/aS}; \Delta)$ will depend on the laser frequency as depicted with the blue dashed curves in the bottom panels of Fig.~3. In particular, the Stokes signal should be strongest for the incident laser blue-tuned from the cavity. This general result for the Stokes scattering is {confirmed by our numerical} calculations performed for various pumping powers ($\Omega^2=10^{-3}$~eV$^2$ to $0.25$~eV$^2$, from red to green lines) and different temperatures ($T=0$~K as dashed lines and 300 K as solid lines),  as shown in the upper panels of Fig.~3(a). 
	
	However, there are also marked differences: we note that when the laser is blue-detuned from the cavity resonances ($\Delta<0$), the profile $S(\omega_{S}; \Delta)$ becomes narrower as we increase the pumping intensity. This behavior results from the buildup of the incoherent population of phonons due to the vibrational pumping and to the dynamical backaction.\cite{aspelmeyer2014cavity} The latter phenomenon, triggered only when the laser is blue-detuned from the cavity resonance ($\Delta < 0$), has been observed in optomechanical systems,\cite{RevModPhys.86.1391} and was recently argued to occur in Raman scattering from molecules.\cite{roelli2014molecular} We discuss this concept in more detail in the following paragraphs.
	
	{Interestingly, the classical Eq.~(\ref{classical}) fails to explain the dependence of the anti-Stokes scattering obtained in the full quantum calculations (Fig.~3(b)). For the weakest driving powers (red lines, $\Omega^2=10^{-3}$~eV$^2$), the $S(\omega_{aS}; \Delta)$ intensity is found to be the largest for the laser frequency $\omega_l$ on resonance with the cavity $\omega_c$ in the absence of thermal pumping of phonons ($T=0$); however this maximum appears as red-tuned from the cavity resonance when the thermal pumping of phonons is produced (at $T=300$~K)}. For stronger driving powers, the intensity plots for $T=0$~K and 300 K start to merge and peak at increasingly blue-shifted frequencies, notably crossing the cavity resonance. As for the supra-linear increase of the Stokes scattering, this surprising property stems from the laser pumping of the vibrational levels. The classical Eq.~(\ref{classical}) for the anti-Stokes $S^{\text{class}}(\omega_{aS}; \Delta)$ intensity does not account for the origin of the vibrations of the molecule, and therefore can only be applied when these are provided by the heated reservoir (it should be noted, however, that a suitable correction to $S(\omega_{aS}; \Delta)$ introducing the vibrational pumping has been proposed by Kneipp \textit{et al.}\cite{PhysRevLett.76.2444}).
	
	To understand the dependencies of the Stokes and anti-Stokes emission on the detuning presented in Fig.~3(a) and (b), we need to consider the vibrational state of the molecule in detail. Since the linearized Hamiltonian cannot be easily solved analytically for an arbitrary detuning $\Delta$, we develop an alternative analysis of the dynamics of the system based on the quantum noise approach,\cite{RevModPhys.86.1391,aspelmeyer2014cavity} which also allows to obtain analytical expressions for the Stokes and anti-Stokes photon emission.
	According to this approach, and similarly to the weak coupling model of a two-level system in a cavity, the rates of relaxation and excitation of vibrations are considered to be proportional to the noise spectrum determined by the characteristics of the plasmonic cavity. This coupling effectively modifies the decay rate of the molecular vibrations to $\gamma_m+\gamma_{\textrm{opt}}$, where the \textit{optomechanical damping} $\gamma_{\textrm{opt}}$ is dependent on the detuning $\Delta$ and the intensity of the incident laser (see Eq.~(\ref{gammaopt})). Furthermore, $\gamma_{\textrm{opt}}$ describes the exchange of energy between the vibrations and fluctuations of the cavity population about its mean value.
	{A description of the quantum noise approach with the explicit expression of $\gamma_{\textrm opt}$ [Eq.~(\ref{gammaopt})] is given in the Methods section. As a result, the intensities of the Stokes and anti-Stokes emissions for general detuning are given within the quantum noise approach by}:
	\begin{equation}\label{qnaS}
	S(\omega_S)\propto \frac{\omega_{S}^4 g^2 n_{a}^{\textrm{coh}} \kappa}{(\Delta+\omega_m)^2+(\kappa/2)^2}  \frac{\neff+1}{\gamma_{\mathrm{opt}} +\gamma_m},
	\end{equation}
	and
	\begin{equation}\label{qnaaS}
	S(\omega_{aS})\propto \frac{\omega_{aS}^4 g^2 n_a^{\textrm{coh}} \kappa}{(\Delta-\omega_m)^2+(\kappa/2)^2} \frac{\neff}{\gamma_{\mathrm{opt}} +\gamma_m},
	\end{equation}
	respectively, where $\neff$ denotes the incoherent phonon population, and $n_a^{\textrm{coh}}$ is the coherent population of the cavity (see the Methods section for the exact definition of these parameters). {The dependence of $\neff$ as a function of detuning and laser intensity, shown in Fig.~3(c) and obtained from application of Eq.~(\ref{ntotal}),  helps to understand the dependences of the Stokes and anti-Stokes emission in Figs.~3(a) and (b). These analytical results are indeed identical to those obtained with the exact numerical solution.}
	
	For the laser blue-tuned from the plasmonic cavity $(\Delta<0)$, the optomechanical damping $\gamma_{\mathrm{opt}}$ becomes negative, { resulting in an increase of the incoherent phonon population $\neff$, as observed in  Fig.~3(c). Consequently, the Stokes and anti-Stokes emission reveal similar dependence on $\Delta$, as shown in Figs.~3(a) and (b).} The Stokes and anti-Stokes peaks' widths exhibit also a narrowing from $\gamma_m$ to $\gamma_m+\gamma_{\textrm{opt}}$\cite{RevModPhys.86.1391} which, interestingly, is compensated for by the denominators in the above equations, ensuring that the integrated intensity of the inelastic scattering is only implicitly dependent on $\gamma_{\mathrm{opt}}$ through $\neff$. Furthermore, we note that {in analogy to other optomechanical systems, we can define a \textit{cooperativity-like parameter} $\tilde{C}=\gamma_{\mathrm{opt}}/\gamma_m$ of the Raman process. For the parameters discussed throughout this work and for blue-tuned laser, $\tilde{C}$ reaches the minimum value of $-0.15$ for $\Delta\approx -0.8\omega_m$ (see Fig. S2 in the Supp. Info.). For $\tilde{C}\leq-1$ (in the case of laser blue-tuned from the cavity $\Delta<0$), the system would exhibit phonon lasing.\cite{roelli2014molecular} }
	
	On the other hand, for red-tuned laser ($\Delta>0$) we enter the {so-called \textit{cooling regime} in optomechanics, in which the vibrations can be suppressed below the thermal population $\nbth$. However, because we are not in the sideband-resolved limit ($\kappa\ll \omega_m$), in a typical SERS configuration} this effect is suppressed, and $\neff$ does not significantly drop below $\nbth$, as illustrated by the solid lines in Fig.~3(c).

	{Last, we focus on an important piece of information regarding the characterization of Raman photon emission that can be obtained from the time- and frequency-resolved correlation\cite{cohentannoudji79a,arXiv_silva14a,hanburybrown56a} between Stokes and anti-Stokes photons emitted from the cavity.} In the steady state, this magnitude can be theoretically calculated through intensity-intensity correlations of the filtered output field,
	\begin{figure}[h]
		\begin{center}
			\includegraphics[width=\columnwidth]{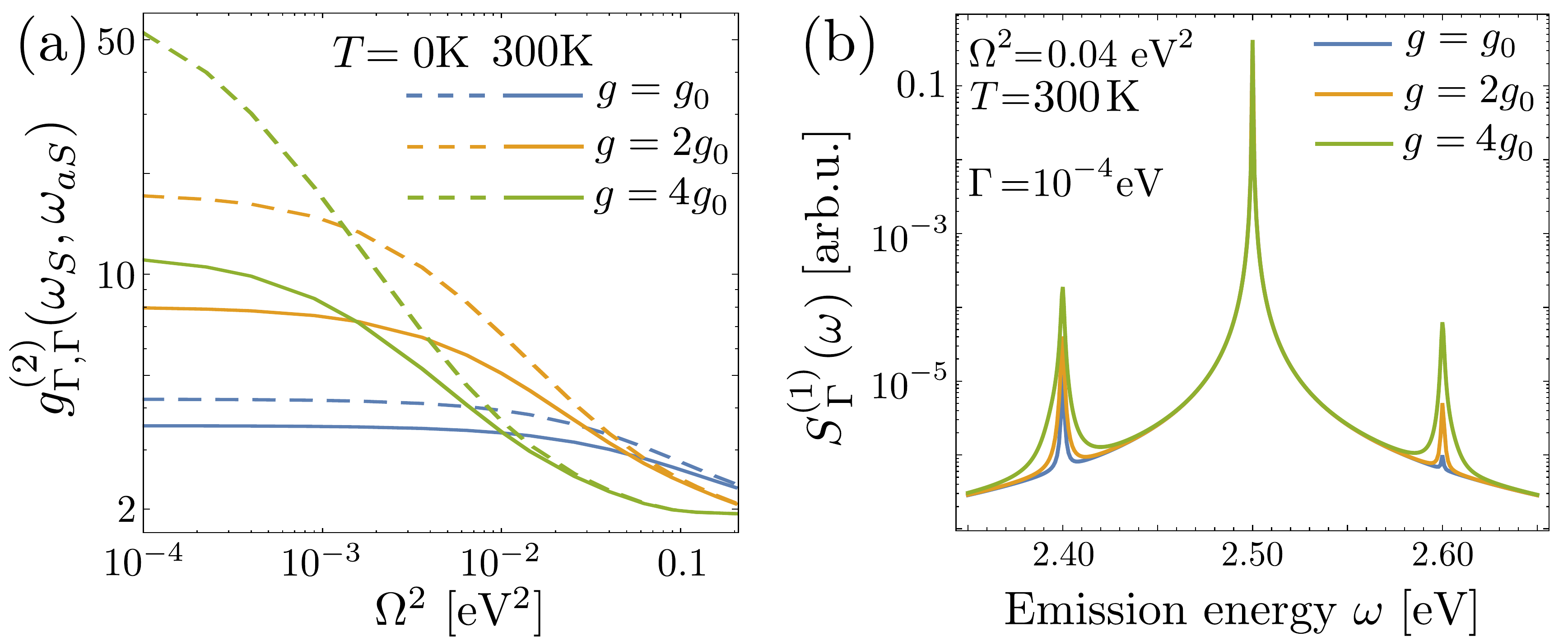}
			\caption{
				(a) Two-photon frequency-resolved Stokes-anti-Stokes correlators $g^{(2)}_{\Gamma \Gamma}(\omega_S,\omega_{aS})$ and (b) physical spectra of emission $S^{(1)}_{\Gamma}(\omega) = \langle \hat{A}^{\dag}_{\omega,\Gamma}(0)\hat{A}_{\omega,\Gamma}(0)\rangle$ calculated for the temperatures $T=0$~K (dashed lines) and $T=300$~K (solid lines), coupling parameters chosen as multiples of $g_0=\kappa/250$: $g=g_0$ (blue lines), $g=2 g_0$ (orange lines), $g=4g_0$ (green lines), and laser tuned to the cavity $\Delta=0$.
			}
			\label{correlations.weak}
		\end{center}
	\end{figure}
	\begin{align}
	\label{eqg2omega2}
	& g_{\Gamma_1, \Gamma_2}^{(2)}(\omega_1,\omega_2;\tau)=\nonumber \\ & \lim_{t \rightarrow \infty}\frac{\mean{:\mathcal{T}[\ud{\hat{A}}_{\omega_1,\Gamma_1}(t)\ud{\hat{A}}_{\omega_2,\Gamma_2}(t+\tau)\hat{A}_{\omega_2,\Gamma_2}(t+\tau)\hat{A}_{\omega_1,\Gamma_1}(t)]:}}{\langle(\ud{\hat{A}}_{\omega_1,\Gamma_1} \hat{A}_{\omega_1,\Gamma_1})(t)\rangle\langle(\ud{\hat{A}}_{\omega_2,\Gamma_2}\hat{A}_{\omega_2,\Gamma_2})(t+\tau)\rangle}\,,
	\end{align}
	{where $\mathcal{T}$ and : denote the time and normal ordering operators, respectively.\cite{Cresser,Knoll:86} Furthermore,}
	${\hat{A}_{\omega_i,\Gamma_i}(t)=\int_{-\infty}^t dt_1
		e^{(i\omega_i-\Gamma_i/2) (t-t_1)} \hat{a}(t_1)}$ is the field detected at frequency~$\omega_i$, within a frequency window~$\Gamma_i$, at time
	$t$. For simplicity we consider $\Gamma_{1}=\Gamma_{2}=\Gamma$ and place the Lorentzian filters at $\omega_{1/2}=\omega_{S/aS}$. The photon correlations $g^{(2)}_{\Gamma, \Gamma}(\omega_S,\omega_{aS};\tau=0)$, calculated using a recently developed method\cite{delvalle12a}, are plotted in Fig.~4(a) as a function of the driving parameter $\Omega^2$ for an environment temperature of $T=0$~K (dashed lines) and $T=300$ K (solid lines). The details of the method of calculation of the correlation function are provided in Supp. Info. The coupling parameters $g$ are considered as multiples of $g_0=\kappa/250$, namely $g=g_0$ (blue lines), $g=2 g_0$ (orange lines), $g=4g_0$ (green lines), with the filter linewidth $\Gamma=0.1$~meV. As shown in Fig.~4(b), for these parameters the physical spectrum of emission $S_\Gamma^{(1)}(\omega)=\langle \hat{A}^{\dag}_{\omega,\Gamma_1}(0)\hat{A}_{\omega,\Gamma_1}(0)\rangle$ is formed by three peaks: the elastic Rayleigh scattering and inelastic Stokes and anti-Stokes contributions (note that $S_\Gamma^{(1)}$ does not include the explicit dependence on the $\omega^4$ factor). 
	We clearly observe in Fig.~4(a) that for weak coherent pumping and in the absence of thermal pumping (dashed lines), the system exhibits strong bunching statistics, which is a signature of strongly correlated emission.\cite{gonzaleztudela13a,ElenaNJP} The physical origin of the strong correlation is {discussed in detail in the Supp. Info. In short,} in the absence of other sources of excitations, the Raman photons are emitted by exchanging a single phonon and are therefore strongly correlated. 
	{This process should exhibit a $g_{\Gamma,\Gamma}^{(2)}(\omega_{S},\omega_{aS})\propto 1/S(\omega_{S})\propto\Omega^{-2}$ dependence, as the rate of coincidences in the detection is determined by the anti-Stokes emission $S(\omega_{aS})$.\cite{kasperczyk2015stokes,jorio} { The results in Fig.~4(a) show this evolution of the correlation for large incident power.  As the thermal pumping (solid lines in Fig.~4(a)) grows,} the anti-Stokes photons are increasingly created through the absorption of phonons which originate from thermal excitation, leading to the quenching of the correlation. Finally,} the dependence of the correlations on the coupling parameter stems from the fact that for small $g$ (and weak anti-Stokes emission), the filters detect primarily the elastically scattered photons (see Fig.~4(b)){, leading to a plateau in $g_{\Gamma,\Gamma}^{(2)}(\omega_{S},\omega_{aS})$ for small $\Omega$}. We note these results are consistent with those recently reported by Kasperczyk \textit{et al.}\cite{kasperczyk2015stokes} from measurements of the Stokes and anti-Stokes pairs emitted from a thin layer of diamond {and by Parra-Murillo \textit{et al.}\cite{jorio} from the theoretical analysis of the Raman scattering}.
	
	\section{Conclusions}
	
	In summary, we have implemented a fully quantum-mechanical model of the inelastic nonresonant Raman scattering from a molecule placed in a lossy plasmonic cavity. Our analytical and exact numerical calculations point to phenomena which go beyond the standard description of Raman scattering such as (i) the onset of stimulated Stokes emission, and (ii) the counterintuitive dependence of the anti-Stokes signal on the detuning of the incident laser from the cavity resulting from the pumping of the molecular vibrations. Furthermore, we demonstrate the strong correlations of Stokes and anti-Stokes pairs emitted from the cavity and analyze the origin of this effect. Finally, we note that our model can be readily applied to a range of physical problems, \textit{i.e.}, investigating the Raman scattering of noncoherent light or the quantum correlations induced in the system.
	
	\section{Methods}
	\subsection{Linearization of the optomechanical Hamiltonian}
	Let us consider the optomechanical Hamiltonian given in Eq.~(\ref{Hamiltonian}). If we decouple the cavity from the molecule (putting $g = 0$) the plasmonic
	system is driven into a steady coherent state with amplitude
	$\alpha_s=\Omega/(\frac{\kappa}{2}+i\Delta)$. The interaction with these
	steady-state plasmons dominates the dynamics of the system in the weak-coupling regime as
	can be seen by the transformation to a displaced basis (with $\ket{\alpha_s}\to\ket{0}$) in the plasmonic system. By, correspondingly,
	replacing $\hat{a}\to \hat{a}+\alpha_s$, we can express the coherent dynamics of the system through the Hamiltonian
	\begin{align}
	\label{eq:Ham0}
	\hat{H} = &\Delta \hat{a}^\dag \hat{a} + \omega_m\hat{b}^\dag \hat{b} - g|\alpha_s|^2(\hat{b}+\hat{b}^\dag)+\nonumber\\ 
	& -g(\alpha_s\hat{a}^\dag +\alpha_s^*\hat{a})(\hat{b}+\hat{b}^\dag) - g \hat{a}^\dag \hat{a}(\hat{b}+\hat{b}^\dag).
	\end{align}
	As we discuss in the Supporting Information in detail, in the regime of parameters used throughout this work, characterized by weak coupling $g$, the last nonlinear term can be neglected, allowing us to write the simplified Hamiltonian
	\begin{equation}
	\label{eq:Ham2}
	\hat{H}' = \Delta \hat{a}^\dag \hat{a} + \omega_m\hat{b}^\dag \hat{b} -
	g|\alpha_s|^2(\hat{b}+\hat{b}^\dag)-g(\alpha_s\hat{a}^\dag +\alpha_s^*\hat{a})(\hat{b}+\hat{b}^\dag),
	\end{equation}
	which yields a purely quadratic/Gaussian dynamics and linear quantum Langevin equations. We will refer to $\hat{H}'$ as a \textit{linearized} Hamiltonian.
	
	In this new Hamiltonian, the coherent driving of the cavity gives rise (\textit{via} the nonlinear interaction) to a linear coupling between phonons and plasmons (the last term in Eq.~(\ref{eq:Ham2})) and an effective coherent driving $g|\alpha_s|^2$ of the phonon mode.
	To remove this driving term, we displace the phononic operators in the same way as for the plasmon, and obtain a new coherent driving of the cavity (and a renormalized detuning), and so successively. We can capture \emph{all} orders of this
	feedback by defining displacements $\alpha_s'$ and $\beta_s'$ through the
	condition that in the displaced basis the Hamiltonian does not contain any linear (\textit{driving}) terms. Such displaced Hamiltonian can be written as
	\begin{equation}
	\label{eq:Ham3}
	\hat{H}'' = \Delta' \hat{a}^\dag \hat{a} + \omega_m\hat{b}^\dag \hat{b}-g[\alpha_s'\hat{a}^\dag +(\alpha_s')^*\hat{a}](\hat{b}+\hat{b}^\dag),
	\end{equation}		
	where displacements $\alpha_s'$ and $\beta_s'$ are defined by
	\begin{equation}\label{alphasP}
	\alpha_s'=\frac{\Omega}{\kappa/2+i\Delta'},
	\end{equation}
	\begin{equation}\label{betasP}
	\beta_s'=\frac{g |\alpha_s'|^2}{\omega_m-i \gamma_m/2},
	\end{equation}
	and the renormalized $\Delta' = \Delta - 2 g \Re (\beta_s')$.
	
	\subsection{Quantum noise approach}
	In the quantum noise approach, the molecular vibrations are coupled to the plasmonic cavity described by the quantum noise density of the plasmon population fluctuations $\hat{a}^{\dag}\hat{a}$ defined as $S_{{\hat{F}\hat{F}}}(\omega)=\int_{-\infty}^{\infty} dt e^{i\omega t}\mean{\hat{F}(t)\hat{F}(0)}$, where $\hat{F}=\hat{a}^{\dag}\hat{a}-\mean{\hat{a}^{\dag}\hat{a}}$ denotes the fluctuations of the plasmon number around the equilibrium, and the average is taken over the state of the uncoupled cavity ($g=0$). From $S_{{\hat{F}\hat{F}}}$ we can obtain the effective relaxation and excitation rates of the vibrations. As a result, the molecular vibrations are in a displaced thermal state. The displacement given by $\beta_s'$, and the incoherent population of the thermal state is 
	\begin{equation}\label{ntotal}
	\neff = \frac{\gamma_m}{\gamma_m+\gamma_{\mathrm{opt}}} \nbth - \frac{\gamma_{\mathrm{opt}}}{\gamma_m+\gamma_{\mathrm{opt}}} \nbrad,
	\end{equation}
	(see the previous section and the Supporting Information for details). The second term in Eq.~(\ref{ntotal}) describes the effective heating of the vibrations by the incoherent population of the cavity (associated with the fluctuations of its population), and dominates the population of phonons in the vibrational pumping regime. It should be noted that this important contribution to $\neff$ cannot be derived by assuming a classical form of the cavity field.\cite{clerk2014basic,aspelmeyer2014cavity} The \textit{optomechanical damping} is defined as
	\begin{align}\label{gammaopt}
	\gamma_{\mathrm{opt}} &= g^2 \left[S_{{\hat{F}\hat{F}}}(\omega_m)-S_{{\hat{F}\hat{F}}}(-\omega_m)\right] = \nonumber\\ 
	&= g^2 n_a^{\text{coh}} \kappa \left\{\frac{1}{(\Delta-\omega_m)^2+(\kappa/2)^2 }-\frac{1}{(\Delta+\omega_m)^2+(\kappa/2)^2 } \right\},
	\end{align}
	with $n_a^{\text{coh}} = |\alpha_s'|^2$ denoting the coherent population of the cavity. Here we simplified the $\hat{F}$ operator by considering the displaced cavity operators $\hat{a}\rightarrow\hat{a}+\alpha_s'$ and dropping the quadratic term: $\hat{F}\approx \sqrt{n_a^{\mathrm{coh}}}(\hat{a}+\hat{a}^\dag)$.\cite{aspelmeyer2014cavity} The first (second) term in the brackets of Eq.~(\ref{gammaopt}) corresponds to the transfer of energy from vibrations to the cavity (from the cavity to the vibrations). The more general form of $\gamma_{\mathrm{opt}}$ includes the \textit{static displacement} of vibrations, which can be accounted for by replacing the detuning $\Delta$ in Eq.~(\ref{gammaopt}) by $\Delta'$ defined earlier. However, in the regime of parameters of interest { in a typical Raman configuration}, this correction is negligible. The second contribution in Eq.~(\ref{ntotal}) - the contribution to the population of vibrations originating from the coupling to the optical cavity - is proportional to $\nbrad$:
	\begin{equation}
	\nbrad = -\frac{(\Delta-\omega_m')^2 + (\kappa/2)^2}{4\Delta \omega_m'}.
	\end{equation}
	Here $\omega_m'$ is the mechanical frequency corrected for the \textit{optical spring effect}.\cite{RevModPhys.86.1391,aspelmeyer2014cavity} This correction however can be neglected for the parameters in the regime of interest {in SERS}.
	
	Finally, we note that in the limit of laser tuned to the cavity resonance $\Delta \rightarrow 0$, there is no DBA mechanism and the optomechanical damping $\gamma_{\textrm{opt}}$ vanishes. In this case, and neglecting the static displacement of vibrations ($\Delta\approx\Delta'$), we reproduce {within the quantum noise approach} the simple expression for the vibrationally pumped incoherent population of vibrations derived earlier from the full quantum-mechanical linearized treatment, {thus making both analytical derivations fully consistent}:
	\begin{equation}
	\lim_{\Delta\rightarrow 0} \neff = \nbth + \frac{4 g_0^2 n_a^{\text{coh}}}{\kappa^2+4\omega_m^2}\frac{\kappa}{\gamma_m} \approx \nbth+s_2 \Omega^2 \frac{\kappa}{\gamma_m}.
	\end{equation}
	Similarly, {the expressions for the photon emission intensities $S(\omega_S)$ and $S(\omega_{aS})$ obtained in Eqs.~(\ref{qnaS}) and (\ref{qnaaS}) with application of the quantum noise approach, also match those obtained from the previous analytical derivation (Eqs.~(\ref{qrtStokes2}) and (\ref{qrtantiStokes2}))}.

	\begin{acknowledgments}
MKS, RE and JA acknowledge funding from the project FIS2013-41184-P of the Spanish Ministry of Economy and Competitiveness, the ETORTEK IE14-393 NANOGUNE'14 project of the Department of Industry of the Government of the Basque Country, project IT756-13 of the Department of Education and Culture of the Basque Country, scholarship AP-2012-4204 from the Spanish Ministry of Education, Culture and Sport and the Fellows Gipuzkoa program of the Gipuzkoako Foru Aldundia through the FEDER funding scheme of the European Union, Una manera de hacer Europa. AGT acknowledges support from the IntraEuropean Fellowship NanoQuIS (625955).
	\end{acknowledgments}
	
	\clearpage
	\onecolumngrid
	\appendix
		\section{Derivation of the parameters}
		\subsection{Cavity-vibrations coupling}
		Here we provide estimates of the coupling parameter $g$ between the Raman transitions in a molecule in vacuum and a plasmonic cavity. We show that this parameter can be related to two magnitudes discussed extensively in the literature: the Purcell factor $F_P$ and the coupling parameter $g_{JC}$ of the resonant emitter-plasmon Jaynes-Cummings Hamiltonian. \cite{Slowik2013,estebanPatch,Hummer,koenderink2010use,fox2006quantum}
		
		The coupling parameter $g$ derived in the main body of the paper is given by
		\begin{equation}\label{def1}
		g=R_{\nu} Q_{\nu}^0 \frac{\omega_c}{\varepsilon_0 V_{\text{eff}}},
		\end{equation}	
		where $R_{\nu}$ is the relevant element of the Raman tensor, $Q_{\nu}^0$ is the zero-point amplitude of the vibrations, and the fraction {on} the right-hand side includes the parameters of the cavity mode - its resonance frequency $\omega_c$ and effective volume $V_{\text{eff}}$. Therefore, we can relate it to the Purcell factor of the mode $F_P$ defined as
		\begin{equation}\label{Pf}
		F_P = \left(\frac{\omega_c}{\varepsilon_0 V_{\text{eff}}}\right) \frac{3 \varepsilon_0}{4 \pi^2 \kappa}\left(\frac{2 \pi c}{\omega_c}\right)^3,
		\end{equation}
		arriving at
		\begin{equation}\label{S3}
		g=R_{\nu} Q_{\nu}^0 F_P \frac{\kappa}{6 \pi \varepsilon_0}\left(\frac{\omega_c}{c}\right)^3.
		\end{equation}
		
		Furthermore, we note that in the derivation of the coupling parameter $g$ we have followed a similar path as for deriving the resonant coupling $g_{JC}$ between a two-level system and a cavity mode in the Jaynes-Cummings model. For the resonant, properly placed and oriented dipolar moment $d$ of the transition of the two-level system, the maximum coupling can be expressed as:\cite{esteban2014strong}
		\begin{equation}\label{S4}
		|g_{JC}| = \frac{d}{\hbar} \sqrt{\frac{\hbar\omega_c}{2 \varepsilon_0 V_{\text{eff}}}},
		\end{equation}
		where $\sqrt{\hbar\omega_c/(2 \varepsilon_0 V_{\text{eff}})}$ yields the electric field as $\hat{E} = i \sqrt{{\frac{\hbar \omega_c}{2 \varepsilon_0 V_{\text{eff}}}}}(\hat{a}-\hat{a}^{\dag})$. Since the Raman processes can be viewed as a two-step process mediated by the virtual state, the coupling $g$ is proportional to the square of $g_{JC}$, as can be shown from Eqs.~(\ref{S3}) and (\ref{S4}):
		\begin{equation}\label{final.g}
		g=\frac{R_{\nu} Q_{\nu}^0}{d^2} 2 \hbar |g_{JC}|^2.
		\end{equation}
		
		We can therefore find an estimate of the absolute value of the coupling coefficient $g$ by recalling some of the reported Purcell factors $F_P$ for cavities resonant in the visible and near-IR regime.\cite{estebanPatch,chen2012metallodielectric,liu2009excitation,Zeng14}
		
		{
			\begin{table}[h!]
				\centering
				\begin{tabular}{| l | l | c | c | c | c | } 
					\hline	
					Reference & Type of cavity & $F_P$ & $\hbar \omega_c$~[eV] & $\hbar \kappa$~[eV] & $\hbar g$ (R6G) [eV]\\
					\hline Esteban \textit{et al.}\cite{estebanPatch} & plasmonic patch & $5\times10^2$ & 0.45 & 0.025 & $2.5 \times 10^{-11}$\\ 
					& antenna &  &  &  & \\ 
					\hline Chen \textit{et al.}\cite{chen2012metallodielectric} & plasmonic particle  & $8\times10^3$ & 1.5 & 0.1 &  $6 \times 10^{-8}$ \\
					& on a dielectric substrate &  &  &  & \\ 
					\hline Liu \textit{et al.}\cite{liu2009excitation} & plasmon mode & $5\times10^3$ & 2.2 & 0.07 & $8\times 10^{-8}$ \\
					& in nanorod &  &  &  & \\  
					\hline Zeng \textit{et al.}\cite{Zeng14} & plasmonic dimer  & $3\times10^3$ & 3.5 & 0.2 & $6 \times 10^{-7}$\\ 
					\hline 
				\end{tabular}
				\caption{Values of the Purcell factors $F_P$ for various types of cavities with resonant energies $\hbar \omega_c$ and widths $\hbar  \kappa$. In the last column we provide the estimates of the coupling coefficient $\hbar g$ calculated for the coupling with the specific transition of a rhodamine 6G molecule (see text below for details).}		
				\label{Stable}
			\end{table}
			
			The values of the elements of the Raman tensor $R_{\nu}$ and the zero-point amplitude $Q_{\nu}^0$ vary significantly for different molecules and each vibrational degree of freedom. The estimates of these values have been provided from theoretical and experimental studies for numerous molecules, including rhodamine 6g (R6G) \cite{watanbe05}
			or various peptides,\cite{gupta09} as well as cluster structures of \textit{e.g.}, silicon.\cite{honea1993raman} To provide estimates of the values of $g$ we consider the specific values of Raman activity of rhodamine 6g. The non-resonant Raman spectra of the R6G molecules exhibit vibrational energies in the range of one hundred of meV and, in the strong-vibrations end of the parameters spectrum, a Raman activity close to $5 \times 10^2~\varepsilon_0^2 \ANGST^4 \text{amu}^{-1}$. In the one-dimensional model of vibrations used throughout this work, {the Raman activity} is equal to $R_{\nu}^2$. Thus, the product $R_{\nu} Q_{\nu}^0$, where $Q_{\nu}^0=\sqrt{\hbar/2\omega_m}$, reaches for the vibrational energy $\hbar \omega_m=0.1$ eV the value of approximately $3\times10^{-30}~\varepsilon_0~m^3$. Including this parameter into Eq.~(\ref{final.g}), we arrive at the upper estimate of the coupling parameter of around $\hbar g\approx 6 \times 10^{-7}$~eV for a molecule placed in a plasmonic dimer nanoantenna.\cite
			{Zeng14}
			
			We should note that, except for the dark plasmon mode in a nanorod,{\cite{liu2009excitation}} the structures listed above are designed to provide large Purcell factors while retaining high quantum efficiency and avoiding quenching of emission from the resonant two-level systems. The latter is of {limited} relevance when designing systems for SERS or TERS (Substrate- or Tip-Enhanced Raman Scattering), since the Raman scattering does not suffer from quenching and, therefore, we can consider other setups, \textit{e.g.} metallic dimers with subnanometer gaps and significantly reduced mode volumes. We can estimate a lower limit for the mode volume from the recent contribution by Barbry \textit{et al.} \cite{fede} where the TDDFT (Time-Dependent Density Functional Theorem) was employed to analyze the response of a dimer of sodium clusters. From the spatial profile of the field localization in the gap, and assuming that all the energy is confined to the gap, we get $V_{\text{min}}\approx 10^{-28}$~m$^3$ for the mode energy of $\hbar \omega_c = 3.3$~eV.
			
			To estimate the coupling parameter expected in such systems, we can apply the definition of the coupling given in Eq.~(\ref{def1}). Taking a more conservative value of the mode volume of $V_{\text{eff}} = 10^{-26}$~m$^3$, and considering the Raman activity of rhodamine 6g, we obtain $\hbar g\approx 1$~meV, which we use throughout this manuscript.
		}
		\subsection{Approximations in the derivation of interaction Hamiltonian}
		
		The complete expression for the interaction Hamiltonian $\hat{H}_I = -\hat{p}_R \hat{E}$, obtained by inserting the expressions for the Raman dipole $\hat{p}_R$ and the electric field $\hat{E}$, is 
		\begin{equation}
		\hat{H}_I = - g \left\{\hat{a}^{\dag} \hat{a} + \frac{1}{2}[-\hat{a}^2 - (\hat{a}^{\dag})^2 + 1]\right\} (\hat{b}^{\dag} + \hat{b}).
		\end{equation}
		Within the rotating wave approximation we remove the first two terms in the square brackets, which yield rapid oscillations at frequencies of $2\omega_c \gg \omega_m$. Furthermore, we can displace the equilibrium position of the vibrations to account for the interaction between vibrations and the vacuum field of the cavity $(-\frac{g}{2}(\hat{b}+\hat{b}^{\dag}))$.
		
		Furthermore, we note that by using the free-space Raman tensor $R_{\nu}$ which describes the interaction of a molecule with an incident planewave in free space, we focus our attention on the \textit{electromagnetic enhancement} mechanism, thus neglecting any contribution from the \textit{chemical enhancement}. 
		
		Finally, for the sake of simplicity, we neglect the unperturbed component of the linear optical polarizability of the molecule. 
		
		\subsection{Coupling of the cavity to the incident light}
		{		
			The coherent driving parameter of the cavity is defined as:\cite{esteban2014strong}
			\begin{equation}\label{OmegaDerive}
			\Omega = \frac{\kappa}{2} \sqrt{\frac{\varepsilon_0 V_{\text{eff}}}{2 \hbar \omega_c}} |E_m^s| = \frac{\kappa}{2} \sqrt{\frac{\varepsilon_0 V_{\text{eff}}}{2 \hbar \omega_c}} K |E_0|,
			\end{equation}
			where $|E_m^s|$ is the maximum of the near field scattered by the structure, which can be introduced through the \textit{field enhancement ability} of the plasmonic system as $K=|E_m^s|/|E_0|$, where $|E_0|$ is the amplitude of the incident coherent illumination. The plasmonic cavity can be also characterized by the Purcell factor $F_P$, the radiative rate enhancement $\Gamma_R/\Gamma_0$ and the quantum yield $\eta$ of a dipolar emitter {placed in such cavity}. Through the reciprocity theorem,\cite{Carminati:98,reciprocity} and for the dipolar mode, these three quantities can be related to $K$ as
			\begin{equation}\label{recip}
			\frac{\Gamma_R}{\Gamma_0} = K^2 = \eta F_P.
			\end{equation}		
			Inserting the definition of the Purcell factor (Eq.~(\ref{Pf})) into the second equality, and plugging the resulting expression for $K$ into Eq.~(\ref{OmegaDerive}), we arrive at
			\begin{equation}\label{Omega}
			\Omega = \frac{1}{4\pi} \sqrt{\frac{3 \varepsilon_0 \lambda_c^3}{2 \hbar}\eta \kappa} |E_0|.
			\end{equation}
			To provide some exemplary values of the parameter $\Omega$, we consider an optical plasmonic cavity with the plasmon frequency $\hbar \omega_c=2.5$~eV, quality factor $Q=10$ ($\hbar \kappa=0.25$~eV) and a conservative quantum yield $\eta=0.01$. Thus, for the strong laser intensity $I=10^{7}$~W/cm$^2$, which yields $|E_0|\approx 8 \times 10^6$~V/m, we get the driving parameter $\hbar \Omega\approx0.11$~eV. Nonetheless, stronger values are necessary to observe clearly the effect of the stimulated Raman scattering in the absence of the dynamical backaction. Therefore, we should consider either a plasmonic cavity providing a stronger enhancement of the incident field, or a stronger illumination source - possibly replacing the continuous illumination source with a pulsed laser (capable of enhancing the electric field by an order of magnitude) with the pulse length larger than the characteristic time of the buildup of phonon population $\propto\gamma_m^{-1}$. Alternatively, the onset of non-linearity could be observed with the current continuous illumination systems by considering stronger coupling with the molecules, either due to the smaller effective volume of the mode, or the stronger intrinsic Raman activity of the probed system (\textit{e.g.} by replacing the R6G with graphene nanosystem or carbon nanotubes). Interestingly, Roelli \textit{et al.}\cite{roelli2014molecular} cite very similar restrictions for the observation of the parametric instability.
		}		
		
		For the strongest laser intensities used throughout the paper, we approach the regime where $\Omega$ and $\omega_c$ become comparable. Nevertheless, even for those strongest pumping coefficients, the pumping term of the Hamiltonian (Eq.~(2) in the main text) is written in the rotating wave approximation. We expect that the inclusion of the counter-rotating terms $i \Omega (\hat{a}e^{-i\omega_l t} + \hat{a}^{\dag}e^{i\omega_l t})$ will not change significantly the dynamics of the system, as they will probably result mainly in energy shifts that can be corrected with appropriate laser detuning.
		
		\section{Linearized Hamiltonian and the analytical treatment of the system}
		In the following section we present an analytical treatment of the system, based on the quantum regression theorem (QRT).\cite{carmichael2009statisticalVol1} To readily apply this method, we first consider a linearized version of our Hamiltonian and verify numerically that this approximation is well-suited for describing the system in the range of parameters discussed here. We then proceed to address the linearized Hamiltonian analytically and present a solution for the case of the incident illumination tuned to the plasmon cavity.
		
		{The response of the system described through the linearized optomechanical Hamiltonian could be alternatively obtained following the \textit{quantum noise approach} {(see the discussion in the manuscript)}.\cite{RevModPhys.82.1155} While this approach is usually applied to derive the limit for the phonon population in the cooling setup,\cite{wilson2008cavity} it can also be extended to describe the effect of the heating of the mechanical degree of freedom.}
		
		\subsection{System}
		Consider the Hamiltonian given by Eq. (1) in the main text in the frame rotating with the frequency of the laser $\omega_l$ ($\Delta=\omega_c-\omega_l$) and with $\hbar=1$:
		\begin{equation}
		\label{eq:Ham}
		\hat{H} = \Delta \hat{a}^\dag \hat{a} + \omega_m \hat{b}^\dag \hat{b}+i\Omega(\hat{a}^\dag-\hat{a})-g\hat{a}^\dag \hat{a}(\hat{b}+\hat{b}^\dag),
		\end{equation}
		With the inclusion of noise fields $a_{\mr{in}}, b_{\mr{in}}$, which are taken to have zero mean value and arise
		from $\delta$-correlated thermal baths,\cite{RevModPhys.86.1391} the dynamics equations for operators $\hat{a}$ and $\hat{b}$ are:
		\begin{align}\label{heisenbergFull}
		\dot{\hat{a}} &=
		-(\kappa/2+i\Delta)\hat{a}+\Omega+ig\hat{a}(\hat{b}+\hat{b}^\dag)+\sqrt{\kappa} a_{\mr{in}}(t),\\
		\dot{\hat{b}} &=-(\gamma_m/2+i\omega_m)\hat{b} +ig\hat{a}^\dag \hat{a}+\sqrt{\gamma_m}b_{\mr{in}}(t).
		\end{align}
		
		\subsection{Linearization procedure}
		
		The Raman scattering setup discussed in this manuscript is characterized by uncommonly large values of the optomechanical coupling parameter. In particular, the \textit{granularity parameter} defined as the ratio $g/\kappa$ is of the order of $10^{-2}$ - such values have so far only been registered in experiments with ultracold atoms.\cite{murch2008observation,RevModPhys.86.1391} The theoretical analysis of the regime where the coupling parameter $g$ and cavity width $\kappa$ are similar\cite{RablPRL11,NunnenkampPRL11} shows the breakdown of this linearization. To relate this result to our work, in Fig. S1 we have analyzed this approximation for the value of the coupling coefficient used {through most of} the manuscript ((a), $g=g_0=\kappa/250$), and significantly increased in (b) ($g= 10 g_0$) and (c)($g=30 g_0$). The strengths of the Stokes and anti-Stokes signals were obtained by solving the master equation with the full (empty squares, Eq.~(\ref{eq:Ham})) or linearized (full circles, 
		Eq.~(10){in the main text}) Hamiltonian. These results indicate that in the regime of parameters discussed in the manuscript both Hamiltonians give identical strengths of the Stokes and anti-Stokes scattering. However, with increasing coupling parameter (by a factor of 10 in Fig. S1(b) or 30 in Fig. S1(c)), the linearized Hamiltonian begins to over-estimate the inelastic scattering.
		\begin{figure}[htbp!]
			\begin{center}\label{SM1}
				\includegraphics[width=.9\columnwidth]{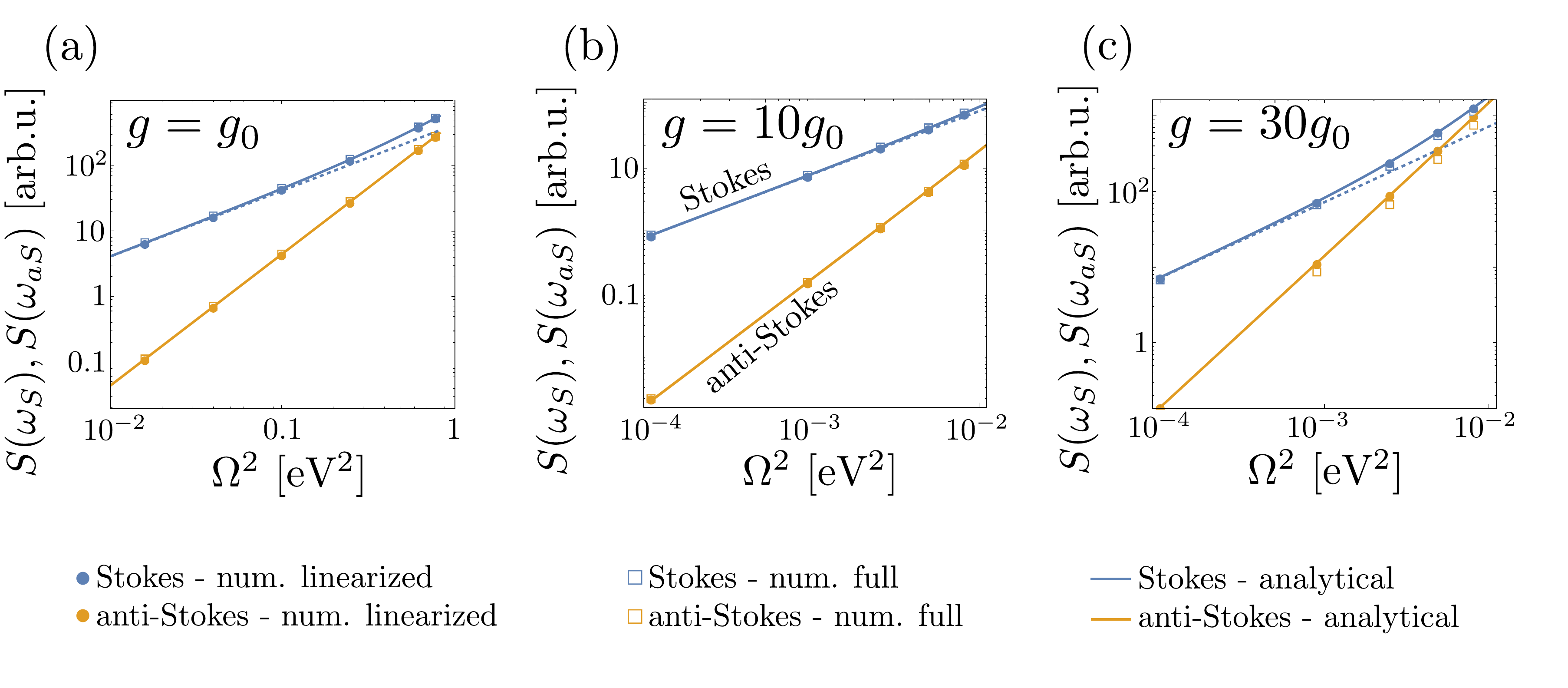}			
				\caption{Dependence of the Stokes (blue) and anti-Stokes ({orange}) scattering on the excitation power and coupling parameter $g$, calculated numerically using full (empty squares, Eq.~(\ref{eq:Ham})) or linearized (full circles, Eq.~(10)) Hamiltonian or analytically (solid lines) from the quantum regression theorem (QRT) using Eqs.(\ref{qrtStokes}) and (\ref{qrtAntiStokes}) for Stokes and anti-Stokes, respectively. Coupling coefficients are given as multiples of the coefficient used throughout the manuscript $g_0=\kappa/250$: (a) $g=g_0$, (b) $g=10g_0$ and (c) $g=30g_0$. The dashed lines denote the linear dependence of the Stokes signal on $\Omega^2$ expected from the classical theories, obtained by taking the first term of Eq.~(\ref{qrtStokes}) exclusively.
				}
			\end{center}
		\end{figure}
		
		In the Methods section, we have introduced the linearized form of the Hamiltonian which does not include any coherent driving terms of the cavity {or} the vibrations. To this goal, we have defined respective coherent amplitudes $\alpha_s'$ and  $\beta_s'$ (see Eqs.~(12) and (13)). Inserting Eq.~(12) into (13) and noting that only the real part $\Re(\beta'_S)$ is relevant, we arrive at a cubic equation, namely
		\begin{equation}\label{sf}
		4g^2[\Re(\beta'_s)]^3-4g \Delta[\Re(\beta'_s)]^2+\left[\left(\frac{\kappa}{2}\right)^2+\Delta^2\right]\Re(\beta'_s)-\frac{g\omega_m\Omega^2}{\omega_m^2+(\gamma_m/2)^2}
		= 0.
		\end{equation}
		The leading order of Eq.~(\ref{sf}) gives the approximations to $\beta_s'$ and $\alpha_s'$, which can be otherwise derived as the coherent amplitudes of the undisplaced operators $\langle \hat{b}\rangle$ and $\langle \hat{a}\rangle$, respectively. The subsequent corrections decrease very quickly and are negligible for our parameters. However, the presence of these terms, and in particular the appearance of higher powers of $\Omega^2$ is evidence of the fundamental \emph{non-linearity} of our system.
		
		
		Finally, we note that displacing the original Liouvillian by $\alpha_s'$ instead of $\alpha_s$  (and neglecting the
		cubic term in the Hamiltonian only then) allows our model to also
		account for higher-order interactions between the plasmons driven by the
		pumping laser and the coherent phonons generated \textit{via} the cubic interaction.	
		
		
		In our numerical calculations, we have implemented the Hamiltonians given by Eq.~(11) with and without the additional non-linear term $\propto \hat{a}^{\dag}\hat{a}$, and solved the corresponding master equation, describing the vibrational and {plasmonic} degrees of freedom in the basis of Fock states. For the strongest pumping and coupling parameters, the calculations were ensured to converge by using up to 15 and 10 Fock states for the description of the vibrational and photonic state, respectively.

		\subsection{Steady state of the linearized Hamiltonian}
		
		From here on, we will focus our attention on the dynamics of system given by the linearized Hamiltonian $H'$ and the master equation (Eq. (2) in the main text). We can thus rewrite the Heisenberg equations given in Eq. (\ref{heisenbergFull}) using the displaced photonic operator $\hat{a}$ as
		\begin{equation}\label{heisenbergLin1}
		\dot{\hat{a}} = -(\kappa/2+i\Delta)\hat{a}+ig\alpha_s(\hat{b}+\hat{b}^\dag)+\sqrt{\kappa}a_{\mr{in}}(t),
		\end{equation}
		\begin{equation}
		\label{heisenbergLin2}\dot{\hat{b}} =-(\gamma_m/2+i\omega_m)\hat{b} +ig(\alpha_s^*\hat{a}+\alpha_s \hat{a}^\dag)+ig|\alpha_s|^2+\sqrt{\gamma_m}b_{\mr{in}}(t).
		\end{equation}
		Denoting by $A$ a column vector with expectation values of operators $(\hat{a}^\dag,\hat{a},\hat{b}^\dag,\hat{b})$ and by $\Gamma_{ab}$ the correlation matrix
		\begin{equation}
		\label{eq:5}
		\Gamma_{ab} = \langle \left( \begin{array}{c}\hat{a}^\dag\\ \hat{a}\\ \hat{b}^\dag \\ \hat{b}
		\end{array}\right)\left(\begin{array}{cccc}
		\hat{a}&\hat{a}^\dag&\hat{b}&\hat{b}^\dag
		\end{array}\right)\rangle,
		\end{equation}
		we can rewrite Eqs. (\ref{heisenbergLin1}) and (\ref{heisenbergLin2}) as
		\begin{align}
		\label{eq:7a}
		\frac{d}{dt}A &= M A +D,\\
		\label{eq:7b}
		\frac{d}{dt}\Gamma_{ab} &=
		M\Gamma_{ab}+\Gamma_{ab}M^\dag+AD^\dag+DA^\dag
		+ E,
		\end{align}
		where 
		\begin{align}
		\label{eq:D}
		D&=ig|\alpha_s|^2(0,0,-1,1)^T,\\
		\label{eq:E}
		E&=\diag([0,\kappa,\gamma_m\nbth,\gamma_m(1+\nbth)]),
		\end{align}
		for $\diag$ denoting a diagonal matrix and $M$ the dynamical matrix
		\begin{align}
		\label{eq:M}
		M &= \left( \begin{array}{cccc}
		-(\kappa/2-i\Delta)&0&-ig\alpha_s^*&-ig\alpha_s^*\\
		0&-(\kappa/2+i\Delta)&ig\alpha_s&ig\alpha_s\\
		-ig\alpha_s&-ig\alpha_s^*&-(\gamma_m/2-i\omega_m)&0\\
		ig\alpha_s&ig\alpha_s^*&0&-(\gamma_m/2+i\omega_m)
		\end{array} \right).
		\end{align}
		Denoting by $\tilde{\Gamma}_{ab}=\Gamma_{ab}-AA^\dag$ the
		covariance matrix, and putting for simplicity $\nbth=0$, we get a simple equation of motion:
		\begin{equation}
		\label{eq:MoM-CM}
		\frac{d}{dt}\tilde{\Gamma}_{ab}=M\tilde{\Gamma}_{ab}+\tilde{\Gamma}_{ab}M^\dag +
		\diag([0,\kappa,0,\gamma_m]). 
		\end{equation}
		
		Computations can be simplified by vectorizing these equations. To this goal, we denote by $\vec{\Gamma}_{ab}$ the column vector formed by the rows of $\tilde{\Gamma}_{ab}$ (and, similarly, by $\vec{E}$ the column vector formed by the rows of $E$) and by $\tilde{M}$ the $16\times 16$ matrix $M\otimes1+1\otimes M^*$. Then Eq. (\ref{eq:MoM-CM}) reads
		\begin{equation}
		\label{eq:meq-cmab2}
		\frac{d}{dt}\vec{\Gamma}_{ab}=\tilde{M}\vec{\Gamma}_{ab}+\vec{E}.
		\end{equation}
		The displacements and covariances at time $t$ are then given by 
		\begin{align}
		\label{eq:6}
		A(t)&=-M^{-1}D+e^{tM}\left( A(0)+ M^{-1} D \right)\\
		\vec{\Gamma}_{ab}(t)&=-\tilde{M}^{-1}\vec{E}+e^{t\tilde{M}}\left(
		\vec{\Gamma}_{ab}(0)+\tilde{M}^{-1} \vec{E} \right), 
		\end{align}
		and thus the steady-state displacements and covariances are
		\begin{align}
		\label{eq:4}
		A_{\mr{ss}} &= -M^{-1}D\\
		\vec{\Gamma}_{ab,\mr{ss}} &= -\tilde{M}^{-1}\vec{E}.
		\end{align}

		\subsection{Quantum regression theorem}
		We can obtain correlation functions by noting that, from the QRT, the equation of motion for $\langle X(t) Y(0) \rangle$ is the same as that for $\langle X(t) \rangle$ and we find for the matrix
		\[C(t)=\langle \left( \begin{array}{c}\hat{a}^\dag(t)\\ \hat{a}(t)\\
		\hat{b}^\dag(t)\\ \hat{b}(t)
		\end{array} \right) \left(
		\begin{array}{cccc}
		\hat{a}(0) & \hat{a}^\dag(0)&\hat{b}(0) & \hat{b}^\dag(0)
		\end{array}  \right)\rangle_{\text{ss}},\]
		where the subscript $\text{ss}$ denotes the steady-state, $\dot{C}(t)=MC(t)+DA^\dag$ and thus 
		\begin{equation}
		\label{eq:3}
		C(t) = -M^{-1}D A^\dag_{\mr{ss}} + e^{tM} \tilde{\Gamma}_{ab,\mr{ss}},
		\end{equation}
		for the steady-state covariance matrix $\tilde{\Gamma}_{ab,\mr{ss}} = \Gamma_{ab,\mr{ss}}- A_{\mr{ss}} A_{\mr{ss}}^\dag$.
		
		Up to the constant terms we then find that the two-time correlator $\langle \hat{a}^\dag(t) \hat{a}(0) \rangle$ used to calculate the spectrum of emission from the cavity is given by the first element of the vector defined as $\exp(t M)$  acting on the first column of the steady-state covariance matrix $\tilde{\Gamma}_{ab,\mr{ss}}$. { We note that the time-dependent contribution to $C(t)$ depends solely on $\tilde{\Gamma}_{ab,\mr{ss}}$, which is independent of the displacement $\alpha_s$. Therefore, the time-dependent part of $\mean{a^\dag(t) a(0)}_{\text{ss}}$ is
			independent of whether we define it through the displaced or un-displaced cavity operators.}
		
		
		For the resonant case ($\Delta=0$) and without thermal pumping
		($T=0$~K) the exponent of $M$ can be found analytically, and we can
		rewrite the above product as a series of expressions with
		exponential factors given by the eigenvalues of $M$: 
		$\{e^{-\kappa/2},e^{-(\gamma_m/2\pm i\omega_m)}\}$. From those, we choose the terms oscillating at frequencies
		$\pm\omega_m$ which decay as {$e^{-\gamma_m t/2}$}, as they govern
		the strengths of Stokes and anti-Stokes scattering. After some
		algebra, we obtain the explicit expressions for the Stokes and
		anti-Stokes emission as 3rd order polynomials of $\Omega^2$. 
		\begin{equation}\label{qrtStokes}
		S(\omega_S) = \frac{4 \omega_S^4}{\gamma_m} (s_2 \Omega^2 + s_4 \Omega^4),
		\end{equation}
		\begin{equation}\label{qrtAntiStokes}
		S(\omega_{aS}) = \frac{4 \omega_{aS}^4}{\gamma_m} a_4 \Omega^4.
		\end{equation}
		The exact formulas for $s_i$ and $a_i$ are somewhat lengthy, but take a simpler form in the limit $\gamma_m\ll\kappa,\omega_m$ and are then given by
		\begin{align}\label{sa}
		s_2 &\approx  \left(\frac{4 g}{\kappa |\kappa-2 i \omega_m|} \right)^2,\\
		s_4 &\approx  \left(\frac{4 g}{\kappa |\kappa-2 i \omega_m|} \right)^4 \frac{\kappa}{\gamma_m} = s_2^2 \frac{\kappa}{\gamma_m},\\
		a_4 &= s_4.					
		\end{align}
		
		The first term in the expression for $S(\omega_S)$, proportional to $\Omega^2$, is dominant for low pumping power. For larger $\Omega$ the second term becomes dominant, and yields the non-linear dependence discussed in the manuscript. Note that the coefficients of the terms proportional to $\Omega^4$ 
		are equal both for Stokes and anti-Stokes, and thus for larger pumping powers the Stokes/anti-Stokes ratio becomes {asymptotically} independent of $\Omega$. 
		
		These two functions are plotted in Fig. S1 with blue (Stokes, $S(\omega_S)$) and orange (anti-Stokes, $S(\omega_{aS})$) lines and match perfectly with the filled circles representing the numerical solution of the linearized Hamiltonian.
		
		If $T>0$, additional terms appear in the expressions for the Stokes and anti-Stokes emission, reflecting (i) the thermal pumping of the molecule, which effectively changes the strength of the anti-Stokes scattering and (ii) the enhancement of the Stokes scattering by stimulated phonon emission. Specifically, we find that Eq.~(\ref{qrtStokes}) can be rewritten
		\begin{equation}\label{qrtStokesT}
		S(\omega_S) =\frac{4\omega_{S}^4}{\gamma_m}[s_2 (1 + \nbth) \Omega^2 + s_4 \Omega^4],
		\end{equation}
		and the anti-Stokes emission becomes
		\begin{equation}\label{qrtAntiStokesT}
		S(\omega_{aS}) = \frac{4\omega_{aS}^4}{\gamma_m} (a_2 \nbth \Omega^2 + a_4 \Omega^4),
		\end{equation}
		where $a_2=s_2$. Thus, in the region of thermal pumping (where we neglect all the terms $\propto \Omega^4$ in the above equations), we recover the well-known formula for the anti-Stokes/Stokes ratio \cite{LeRu20091}
		\begin{equation}\label{ratio}\frac{S(\omega_{aS})}{S(\omega_{S})} \approx \left(\frac{\omega_{aS}}{\omega_{S}}\right)^4 \frac{\nbth}{1+\nbth}.\end{equation}
		
		\subsection{Phonon population}	
		Our analytical approach allows us to write down the explicit expression for the incoherent phonon populations $n_b^{\text{incoh}}$ due to the thermal $\nbth$ and vibrational pumping, and coherent phonon population $n_b^{\text{coh}}$ in the case of the laser tuned to the cavity resonance ($\Delta=0$):
		\begin{equation}
		n_b = \langle \hat{b}^{\dag}\hat{b}\rangle_{\text{ss}} = n_b^{\text{incoh}} + n_b^{\text{coh}},
		\end{equation}
		where 
		\begin{equation}
		n_b^{\text{incoh}} = \nbth + \left(\frac{4 g \Omega}{\kappa}\right)^2  \frac{\kappa+\gamma_m}{\gamma_m|\kappa+\gamma_m+2i\omega_m|^2} \approx \nbth + s_2 \Omega^2 \frac{\kappa}{\gamma_m},
		\end{equation}	
		\begin{equation}
		n_b^{\text{coh}} \approx \left(\frac{2\Omega}{\kappa}\right)^4 \left(\frac{g}{\omega_m}\right)^2.
		\end{equation}
		
		\subsection{Quartic dependence of Stokes intensity on the enhancement of the incident field}			
		By inserting the definitions of $\Omega$ (Eq.~(\ref{OmegaDerive})) and $g$ (Eq.~(\ref{def1})) into the lowest-order expression for $S(\omega_S)$ (Eq.~(\ref{qrtStokes})), we can rewrite $S(\omega_S)\propto K^2/V_{\text{eff}}$, where $K$ is the enhancement of the incident field at the position of the molecule. The inverse volume factor can be, by relating the Purcell factor (Eq.~(\ref{Pf})) to $K^2$ through the reciprocity theorem (see Eq.~(\ref{recip})),\cite{Carminati:98,reciprocity} shown to be proportional to $K^2$, allowing us to recover the expected dependence $S(\omega_S)\propto K^4$.	
		
		It should be also noted that a similar dependence of the anti-Stokes intensity on $K$ can be retrieved only in the thermal pumping regime, where the phonons are primarily provided by the thermal bath, and the anti-Stokes intensity is proportional to $\nbth \Omega^2$. In the vibrational pumping regime, the phonons are provided by the Stokes transitions, and thus we expect to retrieve a higher-order dependence of  $S(\omega_{aS})$ on $K$.
		
		{				
			\subsection{Dynamical backaction and cooperativity}				
			For the laser detuned ($\Delta=\omega_c-\omega_l\neq0$) from the plasmonic cavity, the incoherent population of phonons - both vibrationally and thermally pumped - is further enhanced (for $\Delta<0$) or quenched (for $\Delta>0$) by the onset of the dynamical backaction mechanism.\cite{RevModPhys.86.1391}
			
			In this framework, we have introduced in the text the \textit{cooperativity-like parameter} $\tilde{C}$
			\begin{equation}
			\tilde{C} =\frac{\gamma_{\textrm{opt}}}{\gamma_m}.
			\end{equation}	
			In the limit of $\tilde{C}$ reaching $-1$, we observe the onset of \textit{phonon lasing},\cite{RevModPhys.86.1391,roelli2014molecular} in which the Fock basis required for the description of the final phonon state would be infinitely large. However, in the regime of parameters discussed in our work, $\tilde{C}$ is far from this value. To illustrate this effect, in Fig. S2 we show the values of cooperativity calculated for the pumping intensities used in Fig.~3 in the main text of the manuscript.
			\begin{figure}[htbp!]
				\begin{center}\label{S3fig}
					\includegraphics[width=.5\columnwidth]{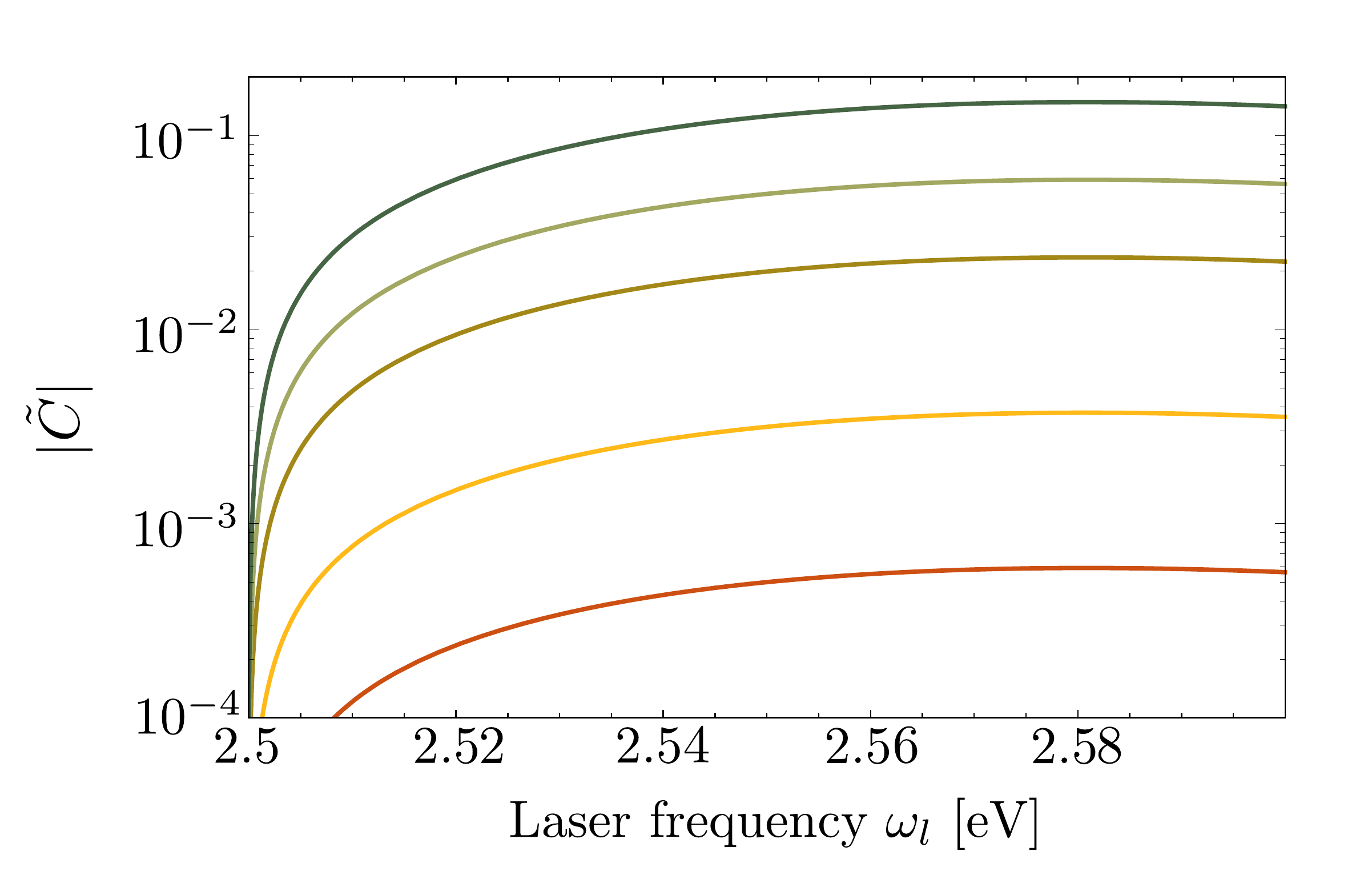}			
					\caption{{Absolute values of $\tilde{C}$ for the pumping intensities $\Omega^2$ from $\Omega^2=10^{-3}$~eV$^2$ (red lines) to $\Omega^2=0.25$~eV$^2$ (green lines). All the parameters of the system are identical to those discussed in the main text of the manuscript.}
					}
				\end{center}
			\end{figure}	
		}				
		
		\section{Threshold for the onset of the phonon-stimulated Raman scattering}
		From Eq.~(3) of the main text of the manuscript, we can derive an approximate criterion for the onset of the phonon-stimulated Raman scattering for the laser tuned to the cavity resonance, defined arbitrarily by the stimulated Stokes emission becoming as strong as the spontaneous emission:
		\begin{equation}
		\nbth + s_2 \Omega^2 \frac{\kappa}{\gamma_m} > 1.
		\end{equation}
		Dropping the first term on the left-hand side, which becomes comparatively very small as the power is increased, and inserting the definitions of $s_2$, $g$ and $\Omega$ from Eqs.~(\ref{def1}), (\ref{sa}) and (\ref{Omega}), we can show that
		{
			\begin{equation}
			s_2 \Omega^2 \frac{\kappa}{\gamma_m} \propto \left(\frac{R_{\nu}Q_{\nu}^0}{V_{\text{eff}}}\right)^2 \frac{\eta \lambda_c}{\gamma_m |\kappa-2i\omega_m|^2} |E_0|^2.
			\end{equation}
			Therefore, the phonon-stimulated processes can become relevant for weaker intensities of the incident laser power if we consider either a molecule with a stronger Raman activity and a weaker decay rate of phonons $\gamma_m$, or a better cavity with smaller effective volume and decay rate $\kappa$ or larger quantum yield $\eta$.				
		}			
		
		\section{Two-photon physical spectrum}
		An important consequence of considering the full quantum model of the SERS is the possibility of studying quantum correlations induced between the phonon and cavity mode. 
		A way of characterizing these correlations may be through quantum state tomography. However, experimentally accessing the intracavity or the phonon field is impossible with the current technology.
		
		An interesting alternative consist of studying the correlations between the Stokes/anti-Stokes photons in the outfield which can be done by inserting filters into the standard Hanbury-Brown-Twiss setup \cite{hanburybrown56a}. Recently, it has been shown how the correlations between different frequencies \cite{delvalle12a,akopian06a, hennessy07a, kaniber08a,
			sallen10a,ulhaq12a} or their full landscapes $(\omega_1,\omega_2)$ \cite{gonzaleztudela13a,ElenaNJP,gonzaleztudela15a} shed light on quantum dynamics that is otherwise hidden in standard spectroscopy or color-blind correlations such as violation of classical Bell and Cauchy-Schwartz inequalities.\cite{sanchezmunoz14b} Moreover, recent advances with streak camera setups allows to do such correlations in the ps scale.\cite{arXiv_silva14a}
		
		Interestingly, the regions of strongly bunched frequency correlations have been linked to the violation of classical inequalities \cite{sanchezmunoz14b} and can be optimized through proper filter engineering \cite{gonzaleztudela15a}, potentially providing access to having a quantum correlated emission from our setup. 
		
		Theoretically, these time and frequency-resolved photon correlations are computed through intensity-intensity correlations:
		\begin{equation}
		g_{\Gamma_1, \Gamma_2}^{(2)}(\omega_1,\omega_2;\tau)=\lim_{t \rightarrow \infty}\frac{\mean{:\mathcal{T}[\ud{\hat{A}}_{\omega_1,\Gamma_1}(t)\ud{\hat{A}}_{\omega_2,\Gamma_2}(t+\tau)\hat{A}_{\omega_2,\Gamma_2}(t+\tau)\hat{A}_{\omega_1,\Gamma_1}(t)]:}_{}}{\langle(\ud{\hat{A}}_{\omega_1,\Gamma_1} \hat{A}_{\omega_1,\Gamma_1})(t)\rangle_{}\langle(\ud{\hat{A}}_{\omega_2,\Gamma_2}\hat{A}_{\omega_2,\Gamma_2})(t+\tau)\rangle_{}},
		\label{eqg2omega2}
		\end{equation}
		where $\hat{A}_{\omega_i,\Gamma_i}(t)=\int_{-\infty}^t dt_1
		e^{(i\omega_i-\Gamma_i/2) (t-t_1)} \hat{a}(t_1)$ is the output field after passing through a Lorentzian frequency filter with central frequency~$\omega_i$, and width $\Gamma_i$, at time $t$. {We note that throughout this section $\hat{a}$ denotes the original operators of the cavity, before displacement.} In principle, in order to compute these correlations one must apply the quantum regression theorem three times and perform the integrals afterwards. However, in this work we use the method recently developed in Ref.~\citenum{delvalle12a}, that avoids this complication by coupling the mode of interest, i.e., $\hat{a}$ in our case, to two-level systems that will play the role of \textit{sensors} (see the schematic in Fig.~S3), with frequencies $\omega_i$ and lifetimes $\Gamma_i$, through the following Hamiltonian:
		\begin{equation}
		\hat{H}_{\text{sens}}=\sum_{i=1,2} \hbar \epsilon_i (\hat{a}^{\dag} \hat{\sigma}^{(i)}_-+\hat{a} \hat{\sigma}^{(i)}_+)\,,
		\end{equation}
		where the coupling $\epsilon_i$ must be sufficiently weak so that the dynamics of the sensors does not perturb the system., i.e.,  $4\epsilon_i^2/\Gamma_i\ll \gamma_s$, where $\gamma_s$ is the smallest transition rate of interest (here $\gamma_s=\gamma_m$). Notice that the following can always be imposed as $\epsilon_i$ is a free non-zero parameter that we can choose at will in our simulation. Under this assumption it can be shown that:
		\begin{equation}
		\label{eq:twophotonspectrum}
		g_{\Gamma_1,\Gamma_2}^{(2)}(\omega_1,\omega_2;\tau)=\lim_{\epsilon_{i}\rightarrow 0}\frac{\mean{:\mathcal{T}[\hat{n}_1(0) \hat{n}_2 (\tau)]:}_{\text{ss}}}{\mean{\hat{n}_1}_{\text{ss}}\mean{\hat{n}_2}_{\text{ss}}},
		\end{equation}
		with $\hat{n}_i=\hat{\sigma}^{(i)}_+\hat{\sigma}^{(i)}_-$. This simplifies the calculation at the cost of a small increase in the dimension of the Hilbert space. For example, in the case of coincidences, i.e., $\tau=0$, that we are mainly interested in, we only need to compute one-time correlators, avoiding the need of the quantum regression theorem.
		
		\begin{figure}[htbp!]
			\begin{center}\label{S2fig}
				\includegraphics[width=.35\columnwidth]{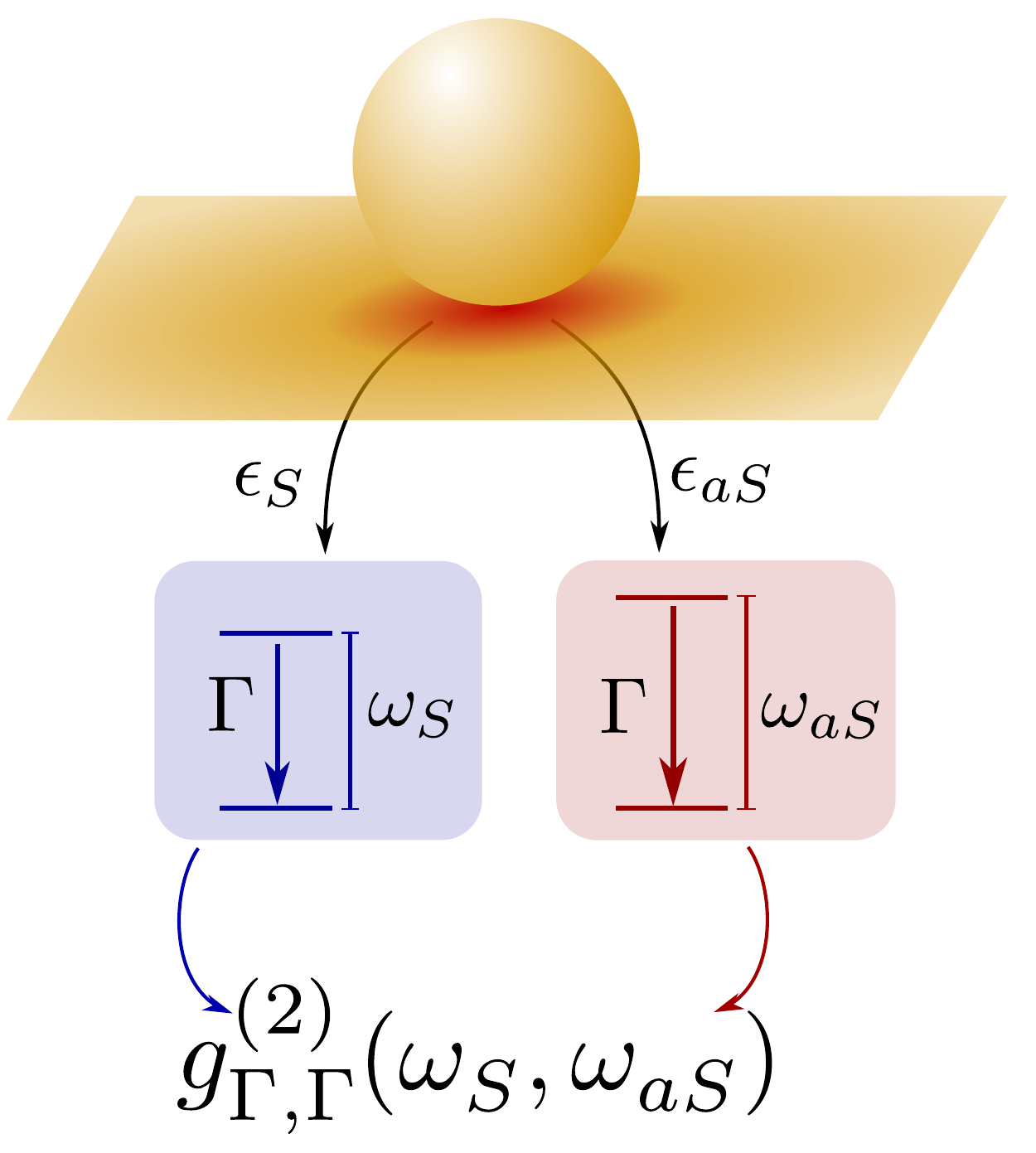}			
				\caption{Two-photon frequency-resolved correlators $g^{(2)}_{\Gamma, \Gamma}(\omega_S,\omega_{aS})$ are calculated by adding two lossy ($\Gamma$) weakly coupled ($\epsilon_S,\epsilon_{aS}\ll \sqrt{\gamma_m \Gamma}/2$) two-level \textit{sensors} and calculating their intensity correlations.\cite{PhysRevLett.109.183601}
				}
			\end{center}
		\end{figure}
		
		
		
		Note that the state of the Stokes/anti-Stokes output modes expected in our
		setting is to good approximation given by a very weak two-mode squeezed
		vacuum state $\ket{\psi}\approx (1+\epsilon \hat{a}^\dag_S \hat{a}^\dag_{aS})\ket{0}$
		reflecting the fact that every anti-Stokes photon must (at
		$T=0$~K) be preceded by a Stokes photon. With the use of $g^{(2)}$ alone this
		quantum-correlated (entangled) state
		is, however, not distinguishable from the classically correlated mixture
		$(1-\epsilon^2)\ket{0}\bra{0} + \epsilon ^2 \ket{2}\bra{2}$, both of which
		show diverging $g^{(2)}$ as $\epsilon\to0_+$.
		
		{				
			\subsection{Dependence of Stokes-anti-Stokes correlations on the parameters of the system}				
			As we show in Fig. 4 in the manuscript, the Stokes-anti-Stokes correlations decrease both with increasing temperature and increasing pumping intensity. Furthermore, the calculated correlations are significantly weaker for the smaller coupling parameter $g$.
			
			\begin{figure}[htbp!]
				\begin{center}
					\includegraphics[width=.9\columnwidth]{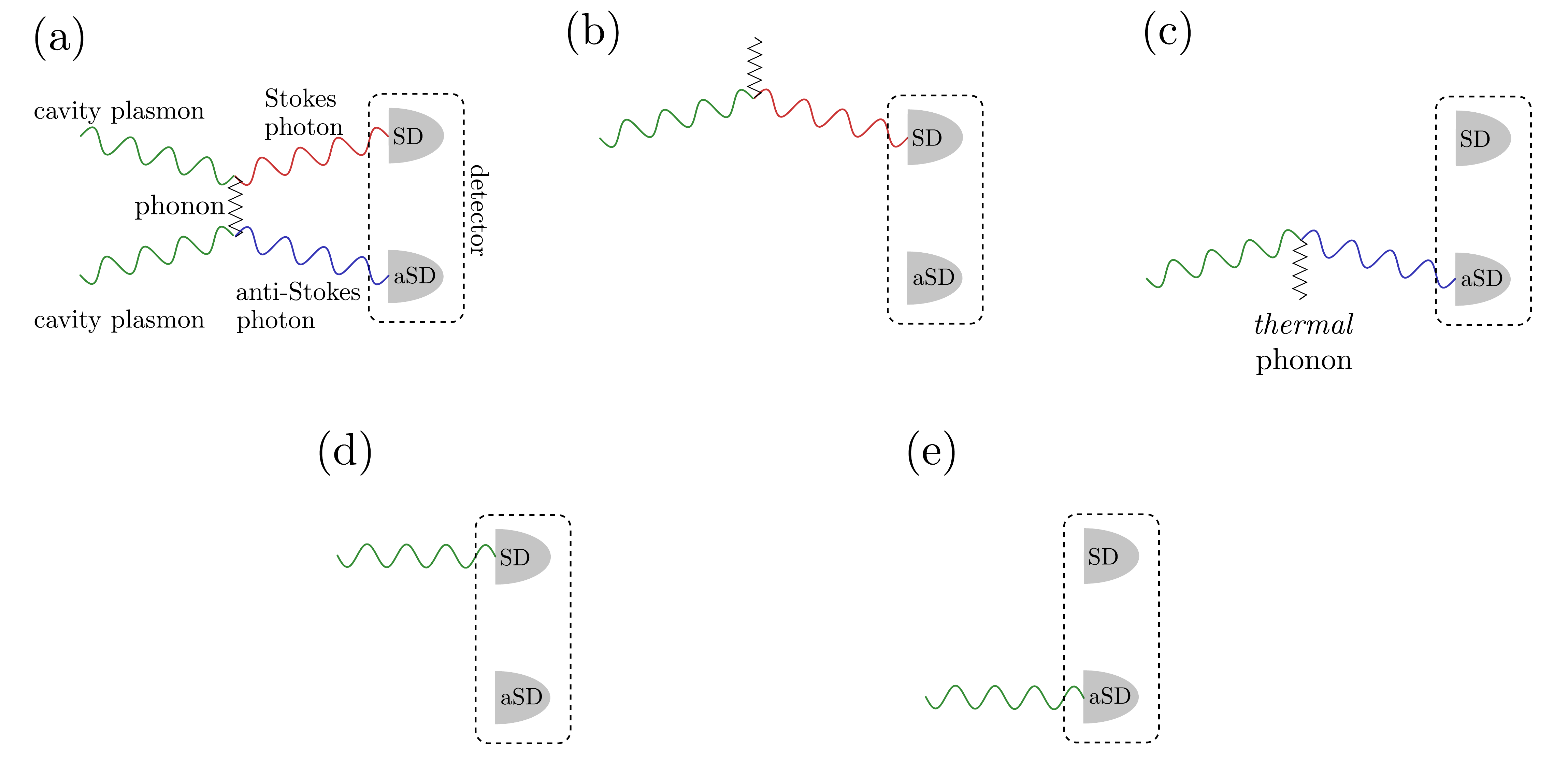}			
					\caption{{Main processes contributing to the correlation
							$g^{(2)}(\omega_S,\omega_{aS},\tau=0)$ between Stokes and anti-Stokes photons
							emitted from a SERS cavity observed by detectors of Stokes (SD) and
							anti-Stokes photons (aSD).
							For weak pumping and zero temperature,
							the dominant process creating anti-Stokes photons is (a) the correlated
							emission of Stokes and anti-Stokes photon generated by exchanging a
							single phonon, while (b) Stokes photons are mainly created by emitting a
							phonon that decays without producing an anti-Stokes transition.
							(c) At non-zero temperatures, the anti-Stokes photons reaching the
							detector can also be generated through the absorption of thermal phonons.
							(d-e) The detectors characterized by a non-zero spectral width $\Gamma$
							can also detect elastically scattered photons instead of the Stokes or
							anti-Stokes photons.}
					}
				\end{center}\label{S4fig}
			\end{figure}	
		}			
		
		To qualitatively describe these dependencies, we consider the Stokes-anti-Stokes correlation $g^{(2)}_{\Gamma, \Gamma}(\omega_S,\omega_{aS},\tau=0)$ defined by the joint probability of detecting a pair of photons, simultaneously reaching the Stokes (SD, see schematics in Fig. S4) and anti-Stokes detectors (aSD), normalized by the probabilities of the SD and aSD detections
		\begin{equation}\label{g2}
		g^{(2)}_{\Gamma, \Gamma}(\omega_S,\omega_{aS},\tau=0) = \frac{P(S,aS)}{P(S)P(aS)}.
		\end{equation}
		Recently, Kasperczyk \textit{et al.}\cite{kasperczyk2015stokes} and Parra-Murillo \textit{et al.},\cite{jorio} have analyzed
		an idealized version of such detection setup, in which the detection
		probabilities $P(S)$ and $P(aS)$ are proportional to the respective
		intensities of the Raman intensities $S(\omega_S)$ and $S(\omega_{aS})$,
		respectively. They showed that if all anti-Stokes photons are produced
		as part of correlated pairs of Stokes and anti-Stokes photons emitted
		from the cavity (as represented pictorially in Fig. S4(a)), then the joint
		probability $P(S,aS)$ should be proportional to the intensity of the
		anti-Stokes emission $P(S,aS)\propto P(aS)$, resulting in
		the correlation $g^{(2)}$ inversely proportional to the probability of
		the Stokes emission ($g^{(2)}\propto1/P(S)\propto
		1/S(\omega_S)\propto\Omega^{-2}$, where the last proportionality is due
		to first order processes such as the one depicted in Fig. S4(b)).
		
		In a realistic scheme, as discussed in the manuscript, this simple
		relationship is modified by the presence of additional processes
		(notably those illustrated in Fig. S4(c)-(e)) that change the dependence on $\Omega$
		of the probabilities appearing in Eq.~(\ref{g2}). 
		\begin{enumerate}
			\item The anti-Stokes detector (aSD) observes photons emitted due to the
			thermal population of phonons ($T > 0$), (see Fig. S4(c)). This effect is
			particularly important for low pumping intensity, where the thermal
			contribution dominates the phonon population.
			
			\item The Lorentzian
			filters used in our detection scheme do not perfectly select the Stokes
			and anti-Stokes photons, breaking the assumption that $P(S)\propto
			S(\omega_S)$ and $P(aS)\propto S(\omega_{aS})$ (see Fig. S4(d) and (e)).
			The importance of this effect can be reduced by using different type of
			detectors, cut-off filters or narrower Lorentzian filters (smaller
			$\Gamma$).
		\end{enumerate}
		
		For sufficiently weak pumping intensity these additional processes
		modify the $\Omega$ dependence of $g^{(2)}$. In particular,
		both processes add a $\propto\Omega^2$ contribution to $P(aS)$ and lead to
		a plateau of the correlations in the weak-pumping regime (see Fig. 4(a)
		in the main text).
		
		Finally, we would like to point that the correlations of emitted light could be potentially modified by the presence of a second, overlapping cavity mode. A similar effect has been reported recently in the study of a bi-modal cavity coupled to a two-level system.\cite{PhysRevLett.108.183601}
		
		\bibliographystyle{apsrev4-1} 
		\bibliography{bibliography}
	
\end{document}